\documentclass[superscriptaddress,twocolumn,amsmath,amssymb]{revtex4-1}% Physical Review B

\usepackage{graphicx}% Include figure files
\usepackage{dcolumn}% Align table columns on decimal point
\usepackage{bm}% bold math
\usepackage{natbib}
\usepackage{amsmath}
\usepackage{amssymb}
\usepackage{float}
\usepackage{datetime}
\usepackage[normalem]{ulem}

\usepackage[usenames,dvipsnames]{xcolor}
\newcommand{\modcol}{\color{black}}

\begin{document}
\author{Francesco Rossella}
\affiliation{NEST, Scuola Normale Superiore 
and Istituto Nanoscienze-CNR, Piazza S. Silvestro 12, I-56124 Pisa, Italy}
\author{Andrea Bertoni}
\affiliation{S3, Istituto Nanoscienze-CNR, 
Via Campi 213a, I-41125 Modena, Italy}
\author{Daniele Ercolani}
\affiliation{NEST, Scuola Normale Superiore 
and Istituto Nanoscienze-CNR, Piazza S. Silvestro 12, I-56124 Pisa, Italy}
\author{Massimo Rontani}
\affiliation{S3, Istituto Nanoscienze-CNR, 
Via Campi 213a, I-41125 Modena, Italy}
\author{Lucia Sorba}
\affiliation{NEST, Scuola Normale Superiore 
and Istituto Nanoscienze-CNR, Piazza S. Silvestro 12, I-56124 Pisa, Italy}
\author{Fabio Beltram}
\affiliation{NEST, Scuola Normale Superiore 
and Istituto Nanoscienze-CNR, Piazza S. Silvestro 12, I-56124 Pisa, Italy}
\author{Stefano Roddaro}
\email{s.roddaro@sns.it}
\affiliation{NEST, Scuola Normale Superiore 
and Istituto Nanoscienze-CNR, Piazza S. Silvestro 12, I-56124 Pisa, Italy}

\title{Nanoscale spin rectifiers controlled by the Stark effect}

\maketitle

{\bf The control of orbital and spin state of single electrons is a key ingredient for quantum information processing~\cite{Loss98,Petta05,Koppens06,Hanson07,Nadj12}, novel detection schemes~\cite{Wabnig09,Giavaras10,Chorley11}, and, more generally, is of much relevance for spintronics ~\cite{Zutic04}. Coulomb~\cite{Averin86} and spin blockade~\cite{Ono02} (SB) in double quantum dots~\cite{vdWiel03} (DQDs) enable advanced single-spin operations that would be available even for room-temperature applications for sufficiently small devices~\cite{Postma01}. To date, however, spin operations in DQDs were observed at sub-Kelvin temperatures, a key reason being that scaling a DQD system while retaining an independent field-effect control on the individual dots is very challenging. Here we show that quantum-confined Stark effect allows an independent addressing of two dots only 5 nm apart with no need for aligned nanometer-size local gating. We thus demonstrate a scalable method to fully control a DQD device, regardless of its physical size. In the present implementation we show InAs/InP nanowire (NW) DQDs that display an experimentally detectable SB up to 10 K. We also report and discuss an unexpected re-entrant SB lifting as a function magnetic-field intensity.}

Pauli exclusion principle and spin conservation in tunnel-coupled multiple quantum dot (QD) systems provide the physical basis of the SB effect~\cite{Ono02}. In suitably-designed solid-state devices this effect can yield an all-electric single-spin manipulation~\cite{Nadj10,Pribiag13,Nowack07,Danon09,SteeleNN09,SteeleNC13}. SB can thus have an important impact on quantum computing architectures~\cite{Loss98} and on single-spin filters and detectors~\cite{Wabnig09,Giavaras10,Chorley11}. Unfortunately, when QD dimensions and relative distance are scaled down to few nanometers or even to the atomic scale, independent control of two QDs is currently viewed as increasingly challenging{\modcol{~\cite{Weber14}}} {\modcol and ultimately plain impossible since it is expected to require fabrication of local gates with different capacitive coupling to the two QDs and thus with dimensions of the order of the inter-dot distance. Here we demonstrate that this is not the case and present an architecture that makes it possible to control the filling of each dot of a DQD system with a single ``large'' split-gate that imposes an electrostatic perturbation uniform at the DQD scale. 

\begin{figure}[ht!]
\begin{center}
\includegraphics[width=0.45\textwidth]{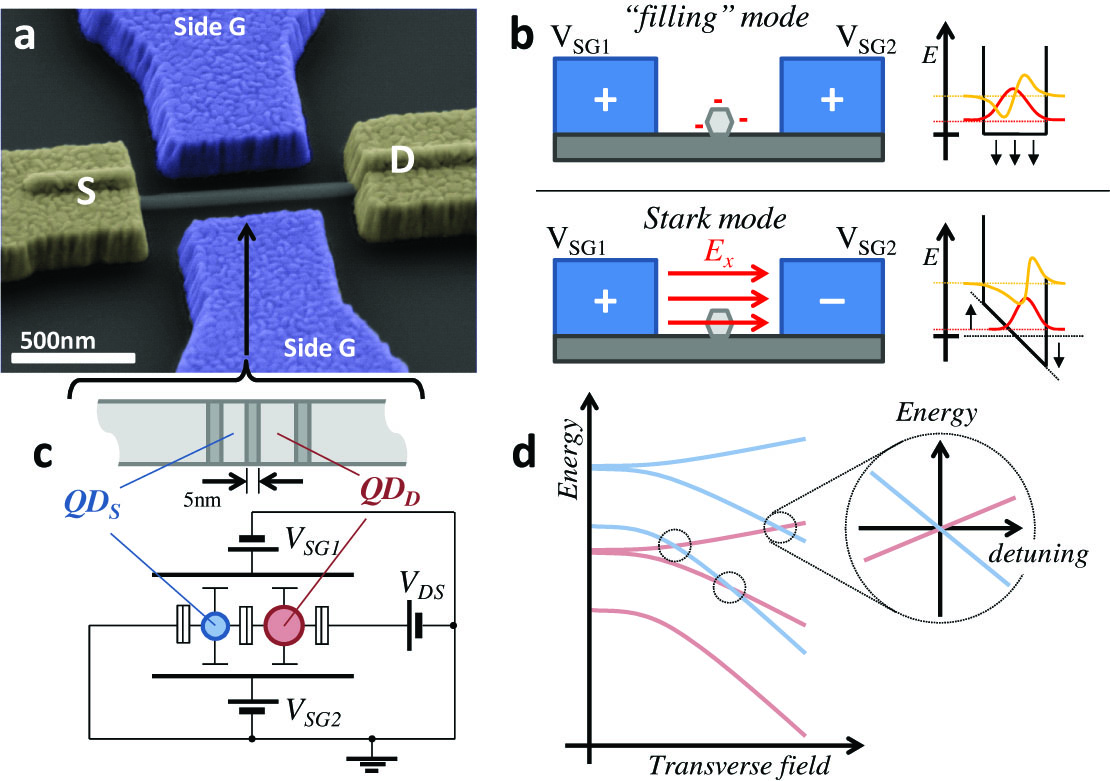}
\caption{{\bf Figure 1. | Stark effect in a double quantum dot.} {\bf a.} Scanning electron micrograph of one of the investigated nanowire-based transistors. A multiple gating technique is used to control quantum states in a one-dimensional InAs/InP heterostructure. {\bf b.} A multigating approach can be used to control both the filling and the transverse mode energies for electrons in the nanowire. {\bf c.} Sketch of the device structure, containing two quantum dots which are subject to the same gating action. {\bf d.} Despite gates are {\modcol two} orders of magnitude larger than the inter-dot distance and are expected to induce a quite uniform electrostatic environment on the active region of the device, Stark effect can easily induce a different energy spectrum evolution in the two QD and provide the equivalent of a detuning parameter in the vicinity of crossings between orbitals. The ideal case depicted in the cartoon corresponds to two radially-identical dots with only a different axial thickness.}
\end{center}
\end{figure}

\begin{figure*}[ht!]
\begin{center}
\includegraphics[width=0.9\textwidth]{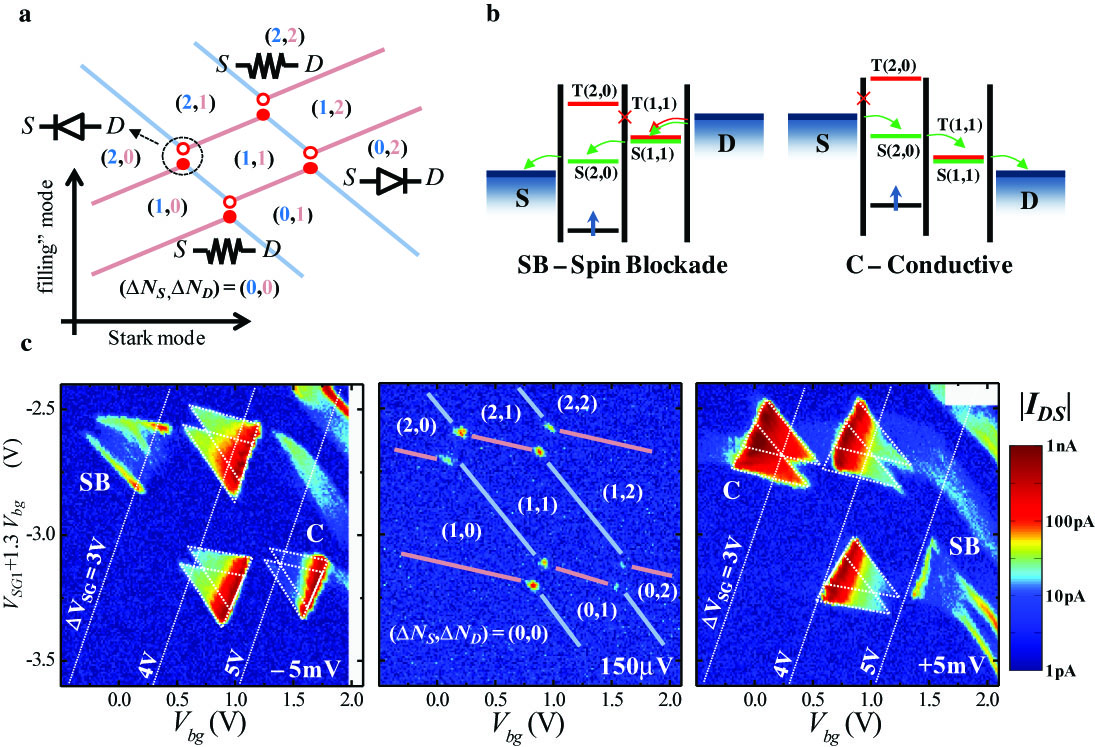}
\caption{{\bf Figure 2. | Tuning spin blockade by Stark effect.} {\bf a.} Stark effect is expected to provide the equivalent of level detuning in a standard DQD architecture, but without the need of any local control of the gating action. Level crossing of Fig.1d will in fact give rise to a honeycomb stability diagram in presence of charging effects. Pink (blue) lines indicate configurations for which the energy of the spin-degenerate level in the S (D) dot falls inside the bias window and configurations that allow sequential tunneling are obtained where such lines cross. Inter-dot capacitive coupling is in this case expected to give rise to two triple points, as show by the red circles~\cite{vdWiel03}. $\Delta N_{S/D}$ indicates the effective filling of the DQD at the level crossing, in excess of electrons occupying filled core orbitals. {\bf b.} In the presence of large orbital energy gaps, the $(1,1)\leftrightarrows (2,0)$ transition can display a strongly non-linear behavior because of Pauli spin blockade. Thus, depending on the specific crossing in panel a, a linear current-voltage response or spin-based rectification can be obtained. {\bf c.} High resolution $|I_{DS}|$ maps in the vicinity of a level crossing in one of the studied devices. Bias triangles at $V_{DS}=\pm 5\,{\rm mV}$ and triple points at low bias ($150\,{\rm \mu V}$) demonstrate a complete control of the DQD operation at $T=1.8\,{\rm K}$. The horizontal and vertical sweep parameters approximately correspond to the Stark mode and ``filling'' mode, as discussed in the main text. Diagonal lines in overlay indicate configuration with an equal value of $\Delta V_{SG}=V_{SG2}-V_{SG1}$, which is proportional to the lateral field $E_x$. A closer inspection of the honeycomb structure reveals unusual non-linearities and an uneven size of the bias triangles. This in not unexpected given the non-standard control mechanism here demonstrated. A clear SB effect is visible in the top-left and bottom-right bias triangles, displaying a rectifying behavior.
}
\end{center}
\end{figure*}

The total filling of a DQD system can be easily controlled using one top/bottom gate, but the selection of {\em which} of the two dots is filled is normally thought to require the ability to raise the potential of one dot with respect to the other. Here we show an alternative and novel approach based on the different Stark shifts that electron levels in two QDs can experience even when subject to the same transverse field.} The key idea is illustrated in Fig.1. In our specific implementation, we demonstrate the first fully-controllable DQD based on heterostructured InAs/InP NWs~\cite{Fuhrer07,Bjork04,Rod11,Romeo12} containing three $5\,{\rm nm}$-thick InP barriers and two InAs sections with a nominal thickness of $20$ and $22.5\,{\rm nm}$. As shown in Fig.1a and further described in Methods, our NW transistors can be gated using the Si/SiO$_2$ substrate and two 600\,{\rm nm}-wide side electrodes. This architecture allows us to control NW population and to impose a tunable electric field in the plane perpendicular to the NW axis (see Fig.1b). When orbitals in distinct QDs cross in energy, Stark effect provides the equivalent of level detuning in a local gated DQD (see sketch in Fig.1c), but does not require the fabrication of field-effect electrodes with a size comparable to that of each QD (see Fig.1d).

Starting from the level crossing of Fig.1d, charging effects are expected to give rise to the honeycomb stability diagram reported in Fig.2a~\cite{vdWiel03} as a function of the ``filling'' and Stark gating modes (see sketch in Fig.1b). Pink (blue) lines indicate configurations corresponding to the alignment of the orbital energy in the S (D) dot with the bias window, which is however still not sufficient to observe transport in the sequential tunnel regime. This becomes only possible at the line crossings, where red circles indicate two split triple point resonances that emerge because of the inter-dot capacitance~\cite{vdWiel03}. The $\Delta N_{S/D}$ overlays indicate the effective DQD filling, on top of eventual occupied core orbitals. When level spacing is large enough to observe quantum Coulomb blockade effects, transport features involving the transition $(1,1)\leftrightarrows(2,0)$ should display rectification due to Pauli SB, as sketched in Fig.2b. It should be noted that in the absence of a good method to address the individual QDs only the ``filling mode'' is available: this implies that in such a case a limited subspace of Fig.2a and a limited device functionality can be accessed~\cite{Fuhrer07}.

The actual experimental demonstration of the DQD control is visible in panel c, where we report a set of colorplots of the current modulus $|I_{DS}|$. The device configuration is controlled by applying a Stark field in the abscissa and by filling the QDs in the ordinate, and exploring the two finite bias $V_{DS}=\pm5\,{\rm mV}$ and the linear regime at $V_{DS}=150\,{\rm \mu eV}$. Since the actual Stark field can only be numerically estimated based on the nominal device geometry, our gate sweeps are designed using the following experimental procedure. In the horizontal direction, the backgate bias $V_{bg}$ is swept while its effect on the DQD charging is balanced by changing the SG1 bias by $\delta V_{SG1}$: here we adopted $\delta V_{SG1}=-\lambda V_{bg}$ where $\lambda=1.3$ was the empirical constant providing the best overall tracking of the device pinch-off in the horizontal scan. In the vertical direction, the DQD was filled by changing a parameter $V_F$ controlling the SG1 bias via the relation $V_{SG1}=V_F-\lambda V_{bg}$. The SG2 electrode can provide a further control but was left grounded in the measurement we report here. The corresponding Stark field in the device plane ($E_x$) can be estimated using the simulations reported in the Supplementary Information and diagonal lines in overlay, indicating the lateral imbalance $\Delta V_{SG}=V_{SG2}-V_{SG1}$. Importantly, $E_x$ is found to be of the right order of magnitude and configurations with an equal lateral field are compatible with the experimental DQD detuning direction. We note that the stability diagram displays a non-linear dependence of dot filling as a function of gating parameters and, for instance, the bias triangle at the bottom-right end of Fig.2c is smaller that its SB-counterpart in the top-left corner. Since the orbitals mediating transport at the visible QD resonances are the same, this effect is surprising if lever arms are interpreted in terms of classic capacitive coupling. In our case, however, Stark effect is expected to have an impact too and to lead to a non-linear evolution of the level energies as a function of the gate voltages.

\begin{figure}[ht!]
\begin{center}
\includegraphics[width=0.45\textwidth]{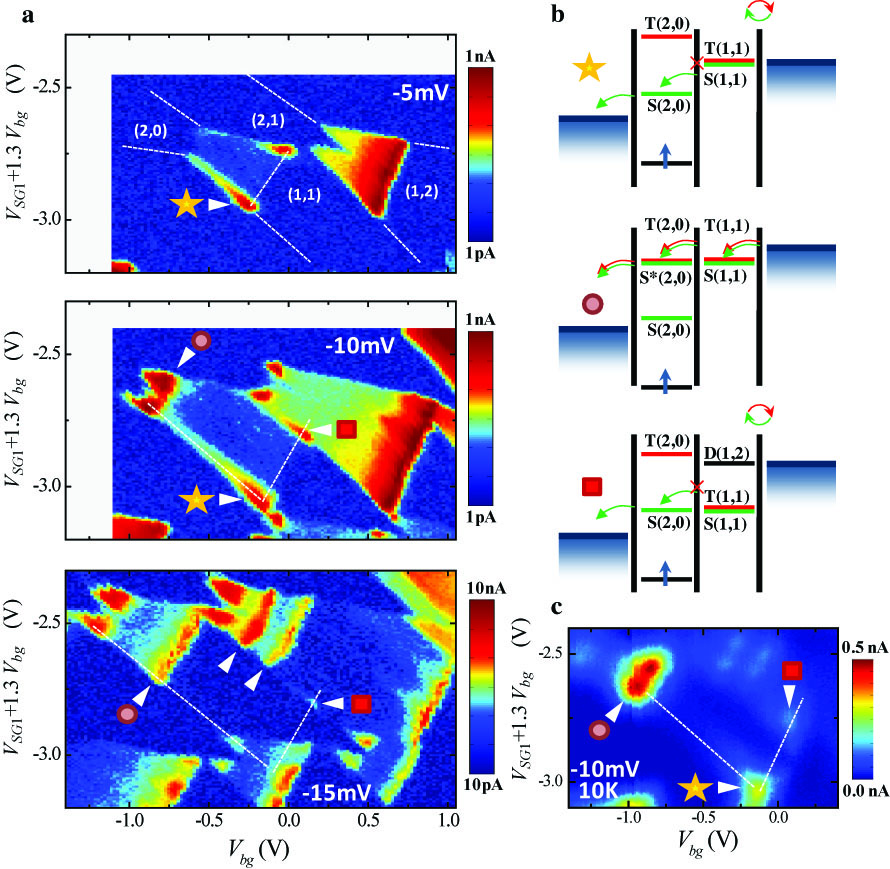}
\caption{{\bf Figure 3. | Finite-bias breakdown of spin blockade.} {\bf a.} Spin-blockade triangles $(1,1)\rightarrow(2,0)$ for different finite bias $V_{DS}=-5$, $-10$ and $-15\,{\rm mV}$ and $T=1.8\,{\rm K}$. The edge of the spin-blockaded region (marked by a golden star) is always conductive since the energy alignment between the right dot and the right lead provides an efficient spin relaxation mechanism. The first non-blocking excited state occurs at an energy $\Delta_{ST}=9.1\pm0.3\,{\rm meV}$: it becomes visible for $|eV_{DS}|>\Delta_{ST}$ and is marked by a purple circle. A further blockade lifting mechanism is also visible at large bias when a further filling configuration $ {\modcol (1,2)}$ becomes available within the biasing window of the spin-filtering transition $(1,1)\rightarrow(2,0)$. This happens when different bias triangles {\em overlap}. The red square indicates level configurations allowing a spin relaxation of the right dot spin via a QD-lead resonance $(1,1)\leftrightarrows{\modcol(1,2)}$. {\bf b.} Sketch of the DQD configurations corresponding to the breakdown of spin blockade visible highlighted in the experimental data in panel a. {\bf c.} Spin blockade detection at $10{\rm K}$ and $V_{DS}=-10\,{\rm mV}$: current suppression at the bias triangle base is still well-evident even at this relatively large temperature.}
\end{center}
\end{figure}

Results of Fig.2 provide the first successful demonstration of a SB rectifier which can be electrically reconfigured by using Stark effect induced by a uniform transverse field rather than standard capacitive lever arms and local gates. Importantly, the method we describe does not require any fine alignments nor nanometer scale gate electrodes and is thus particularly relevant in view of a further downscale {\modcol of} this kind of architecture. The performance and characteristics of our specific physical implementation of a Stark-controlled spin rectifier are further reviewed in Fig.3a, where we report the $(1,1)\to(2,0)$ and $(1,2)\to(2,1)$ bias triangles for three different negative applied bias values: $V_{DS}=-5$, $-10$ and $-15\,{\rm mV}$. A rather large singlet-triplet gap $\Delta_{ST}=9.1\pm0.3\,{\rm meV}$ marks the excited-state lifting of the SB at large bias (purple circle). Two additional SB-lifting mechanisms are sketched Fig.3b. One involves the resonance ${\modcol(1,0)}\leftrightarrows(1,1)$ between the drain-side QD ($QD_D$) and the adjacent lead (gold star). A so-far undocumented behavior can be observed at the ``collision'' with the other bias triangle pairs (red square): this implies a spin substitution via a resonance $(1,1)\leftrightarrows{\modcol(1,2)}$. The very compact multidot system realized here gives rise to rather large energy-level separations in the two QDs and these make it possible to observe SB effects up to unusually high temperatures (see data in Fig. 3c taken at $T\approx 10\,{\rm K}$). Further details about SB and excited states are reported in the Supplementary Information.

\begin{figure}[ht!]
\begin{center}
\includegraphics[width=0.45\textwidth]{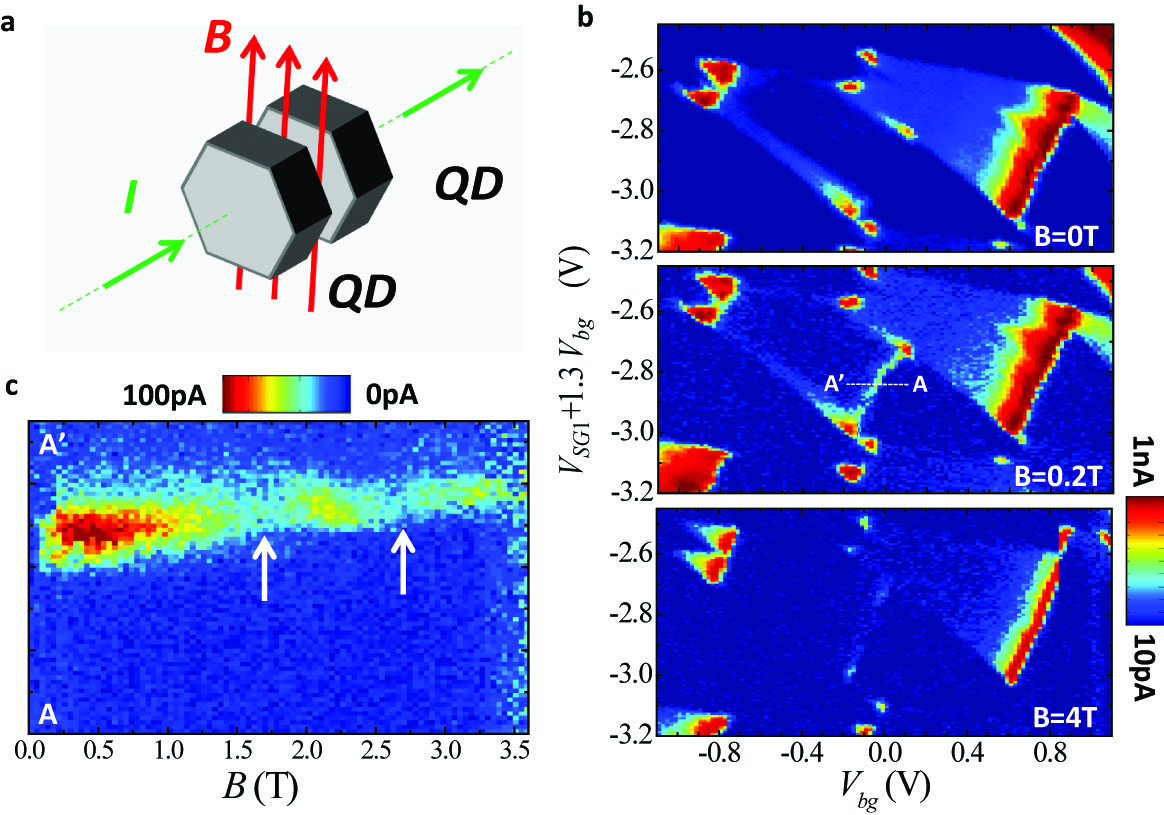}
\caption{{\bf Figure 4. | Re-entrant magnetic lifting of the spin blockade.} {\bf a.} Magnetic field configuration during the measurements. {\bf b.} Bias triangles in SB configuration $(1,1)\to(2.0)$ at $B=0$, $0.2$ and $4\,{\rm T}$ and for $V_{DS}=-10\,{\rm mV}$. A small magnetic field perpendicular to the NW is sufficient to lift the spin blockade, consistently with previous results. At larger magnetic fields the overall conduction in the DQD system tends to be suppressed. {\bf c.} A more subtle evolution is visible when sweeping $B$ between $-3.6\,{\rm T}$ and $+3.6\,{\rm T}$, as a function of the detuning (data taken along the white segment crossing the base of the triangle, as visible in the left panels). The experimental curve is symmetric in $B$ and its average is plotted in the panel: an unexpected oscillatory pattern is observed in the spin relaxation with re-entrant SB marked by arrows and leakage peaks in the $0.2-1.2\,{\rm T}$ range and at $B=2.2$ and $3.3\,{\rm T}$. A likely explanation entails a coupling modulation caused by {\modcol a magnetic field flux through} the finite-size tunnel barrier separating the two dots.}
\end{center}
\end{figure}

We investigated also the magneto-conductance of the DQD system and observed a novel SB lifting phenomenology in the presence of intense magnetic fields applied perpendicularly to the NW axis (see Fig. 4a). Figure 4b shows the recorded SB-bias triangle evolution from zero magnetic field (top panel), to a small field of $200\,{\rm mT}$ (middle panel), to an intense one ($4\,{\rm T}$, bottom panel). Consistently with the established phenomenology for SB in the strong inter-dot coupling regime, SB is maximum at $B=0\,{\rm T}$ and is then lifted by a small magnetic field: this is attributed to spin-phonon coupling mediated by spin-orbit interaction~\cite{Pfund07,Golovach04}. Differently from what {\modcol is} typically reported, however, a re-entrant oscillatory evolution of the SB is observed at higher fields: the effect is highlighted in the cross-section scan in Fig.4c. Despite the rather large excitation energies $\approx 10\,{\rm meV}$ of both $(2,0)$ and $(0,2)$ configurations (see Supplementary Information), the singlet and triplet $(1,1)$ configurations are expected to be close in energy and to be sensitive to inter-dot coupling, since their exchange-energy splitting is proportional to the square of the tunneling amplitude~\cite{Loss98,Bellucci2004}. We thus speculate that the origin of this re-entrant SB is the oscillation of inter-dot coupling with the field, which in turn modulates the exchange energy and hence singlet-triplet mixing. The latter is induced by spin-orbit and hyperfine interactions, which are both expected to be sizable in InAs. {\modcol The general decrease of the conductivity of the InP barriers at very large fields can be ascribed to known magnetotunneling phenomenology~\cite{Vdovin2000}.}

\begin{figure}[ht!]
\begin{center}
\includegraphics[width=0.45\textwidth]{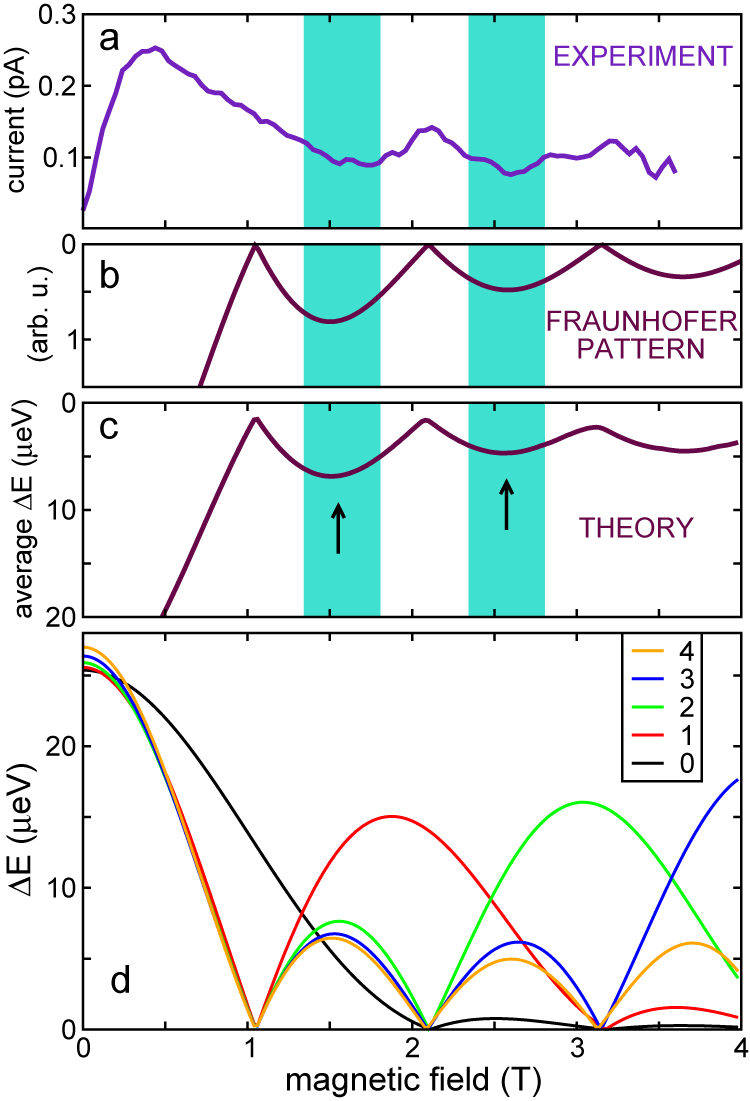}
\caption{{\bf Figure 5. | Coupling oscillation in finite-size tunneling.} 
{\modcol {\bf a.} Crosscut of the current map of Fig.4b taken along the current peak crest, averaged over a fixed bias width. {\bf b.} Absolute value of Fraunhofer diffraction pattern. The diffraction angle is rescaled to match the measured location of current minima.} {\bf c.} Calculated average bonding-antibonding energy splitting $\Delta E$ of the lowest ten radial orbitals of a model double-dot system vs $B$ (the {\modcol first five} single-orbital splittings are plotted in the following panel). The arrows and shaded regions point to the matching between this plot and the measured current. {\modcol We note an excellent agreement between the Fraunhofer pattern and the re-entrant lifting of SB}. {\bf d.} Numerical simulation of the bonding-antibonding energy splitting $\Delta E$ of the five lowest radial orbitals in a double-dot system. Aharonov-Bohm-oscillations are always observed and several zero locations are shared by different orbitals.}
\end{center}
\end{figure}

Figure 5a shows the cross-cut of the colorplot of Fig.4c along the current peak crest, averaged over a fixed bias width. The location of valleys {\modcol compares well with} the sequence of maxima of a Fraunhofer pattern, whose modulus is displayed in Fig.5b after rescaling the diffraction angle {\modcol (the only free parameter) and flipping the ordinate to account for the inverse correlation between tunnel coupling and SB leakage in the explored regime~\cite{Pfund07}. Such matching is indeed consistent with our device architecture: heterostructured barriers and the sharply-defined radial extension of the orbitals implement a spatially-extended tunnel geometry, which is in general expected to lead to a Frauhofer-like modulation of the coupling as a function of the magnetic flux piercing the barrier}. This hypothesis could find a nice confirmation in numerical simulations. In Fig.5d we report the bonding-antibonding splitting $\Delta$E of the lowest-lying radial orbitals of a simplified two-dimensional model of coupled dots with a witdh of $140\,{\rm nm}$, a barrier of $5\,{\rm nm}$ and a dot thickness of $22.5\,{\rm nm}$. Clearly, the {\modcol magnetic field} oscillation of the tunneling amplitude $\Delta E$ is generic to all orbitals, with many sharing the same locations of zeros. Since we ignore the exact filling of our system as well as the effective symmetry of the radial confinement potential, likely prone to different sources of disorder, in Fig.5c we average $\Delta E$ over the first ten orbitals (see Supplementary Information for details of the numerical simulations and the inclusion of a finite temperature). The outcome is very similar to the Fraunhofer patter{\modcol{n}} of Fig.5{\modcol{b}} and compares well with the measured current oscillations (see arrows and shaded regions in Fig.5 and corresponding marks on Fig.4c). At low field, the computed $\Delta E$ is very large and the measured current is indeed vanishes, however, the exact SB lifting profile is dominated by a more complex phenomenology as already reported~\cite{Pfund07,Golovach04}.
 
In conclusion, we have demonstrated a novel method to control spin rectification in a DQD that requires no local/aligned nanogates, which represent a roadblock to scaling. We provided for the first time a demonstration of full-control of a device based on heterostructured InAs/InP NWs and, thanks to the large quantum and electrostatic energy gaps obtained, a clean SB effect up to $10\,{\rm K}$. A novel re-entrant SB lifting as a function of the magnetic field was highlighted. Present results and analysis indicate this is consistent with Fraunhofer-like interference effects on a finite-size tunnel barrier.

{\bf Methods.} 

Devices were built starting from InP/InAs NWs grown by metal-assisted chemical beam epitaxy in a Riber C-21 reactor, using tributyl-arsine (TBAs, line pressure $1\,{\rm Torr}$), trimethyl-indium (TMIn, line pressure $0.7\,{\rm Torr}$) and tributyl-phosphate (TBP, line pressure $1\,{\rm Torr}$) as the metallorganic precursors~\cite{Rod11,Romeo12, Panda2012}. The metallic seed for the growth was obtained by thermal dewetting of a thin Au layer evaporated on InAs (111). All measurements reported refer to devices built starting from a NW with a diameter of 70 nm and a total length of almost $2\,{\rm \mu m}$, based on SEM imaging performed at the end of the measurement session. The nanostructures were transferred by drop-casting to a SiO$_2$/Si substrate and contacted by a Ti/Au ($10/150\,{\rm nm}$) metallic bilayer. Prior to the metal deposition, the contact regions were exposed to an NH$_4$S$_x$ solution in order to passivate the surface and minimize the effect of the surface oxide. The final device could be controlled by gating using top lateral electrodes fabricated in parallel to the ohmic contact as well as the Si substrate, which was heavily doped and had a nominal resistivity of $0.001-0.005\,{\rm \Omega\cdot cm}$. The numerical simulations to compute confined quantum states were performed by an exact diagonalization of the single-particle Schr\"odinger equation discretized with the finite volume method. The magnetic field was included by Peierls substitutions in the Landau gauge (see Supplementary Information). Coulomb blockade measurements where performed in a closed-cycle He3 system at various temperatures ranging from a base of $250\,{\rm mK}$ up to a top temperature of about $15\,{\rm K}$. The control of SB in the DQD by quantum confined Stark effect was demonstrated on three different devices.

{\bf Acknowledgements.}

M.R. acknowledges funding from FP7 Marie Curie ITN INDEX and MIUR-PRIN2012 MEMO. F.R. acknowledges the partial financial support of the MIUR through the FIRB project RBFR13NEA4 ``UltraNano''. We thank Elisa Molinari for her supportive effort and Elena Husanu for STEM imaging. 

{\bf Author contributions.}

S.R. conceived and designed the experiment. D.E. and L.S. grew the nanowires and fabricated the devices. F.R. and D.E. performed the experiment. F.R. S.R., A.B. and M.R. analyzed the data. All the authors contributed to the writing and discussion of the manuscript.


\begin{thebibliography}{99}
%==============================================================================
% [1] citato x QC all'inizio
% ``Quantum computation with quantum dots''
\bibitem{Loss98} 
D. Loss, 
D.~P. DiVincenzo, 
{\em Quantum computation with quantum dots}, 
Phys. Rev. A {\bf 57}, 120-126 (1998).
%==============================================================================
% [2] citato x QC all'inizio
% ``Coherent manipulation of coupled electron spins in semiconductor quantum dots''
\bibitem{Petta05} 
J.~R. Petta, 
A.~C. Johnson, 
J.~M. Taylor, 
E.~A. Laird, 
A. Yacoby, 
M.~D. Lukin, 
C.~M. Marcus, 
M.~P. Hanson, 
A.~C. Gossard, 
{\em Coherent manipulation of coupled electron spins in semiconductor quantum dots}, 
Science {\bf 309}, 2180-2184 (2005).
%==============================================================================
% [3] citato x QC all'inizio
% ``Driven coherent oscillations of a single electron spin in a quantum dot''
\bibitem{Koppens06} 
F.~H.~L. Koppens, 
C. Buizert, 
K.~J. Tielrooij, 
I.~T. Vink, 
K.~C. Nowack, 
T. Meunier, 
L.~P. Kouwenhoven, 
L.~M.~K. Vandersypen, 
{\em Driven coherent oscillations of a single electron spin in a quantum dot}, 
Nature {\bf 442}, 766-771 (2006).
%==============================================================================
% [4] citato x QC all'inizio
% ``Spins in few-electron quantum dots''
\bibitem{Hanson07} 
R. Hanson, 
L.~P. Kouwenhoven, 
J.~R. Petta, 
S. Tarucha, 
L.~M.~K. Vandersypen, 
{\em Spins in few-electron quantum dots}, 
Rev. Mod. Phys. {\bf 79}, 1217-1265 (2007).
%==============================================================================
% [5] citato x QC all'inizio
% ``Spectroscopy of Spin-Orbit Quantum Bits in Indium Antimonide Nanowires''
\bibitem{Nadj12}  
S. Nadj-Perge, 
V.~S. Pribiag, 
J.~W.~G. van den Berg, 
K. Zuo, 
S.~R. Plissard, 
E.~P.~A.~M. Bakkers, 
S.~M. Frolov, 
L.~P. Kouwenhoven, 
{\em Spectroscopy of Spin-Orbit Quantum Bits in Indium Antimonide Nanowires}, 
Phys. Rev. Lett. {\bf 108}, 166801 (2012). % page numbers do not exist
%==============================================================================
% [6] citato all'inizio, detection
% ``A quantum dot single spin meter''
\bibitem{Wabnig09} 
J. Wabnig, 
B.~W. Lovett, 
{\em A quantum dot single spin meter}, 
New J. Phys. {\bf 11}, 043031 (2009). % page numbers do not exist
%==============================================================================
% [7] citato all'inizio, detection
% ``Spin detection at elevated temperatures using a driven double quantum dot''
\bibitem{Giavaras10} 
G. Giavaras, 
J. Wabnig, 
B.~W. Lovett, 
J.~H. Jefferson, 
G.~A.~D. Briggs, 
{\em Spin detection at elevated temperatures using a driven double quantum dot}, 
Phys. Rev. B {\bf 82}, 085410 (2010). % page numbers do not exist
%==============================================================================
% [8] citato all'inizio, detection
% ``Transport Spectroscopy of an Impurity Spin in a Carbon Nanotube Double Quantum Dot''
\bibitem{Chorley11} 
S.~J. Chorley, 
G. Giavaras, 
J. Wabnig, 
G.~A.~C. Jones, 
C.~G. Smith, 
G.~A.~D. Briggs, 
M.~R. Buitelaar, 
{\em Transport Spectroscopy of an Impurity Spin in a Carbon Nanotube Double Quantum Dot}, 
Phys. Rev. Lett. {\bf 106}, 206801 (2011). % page numbers do not exist
%==============================================================================
% [9] MUST - citato all'inizio, spintronics
% ``Spintronics: Fundamentals and applications''
\bibitem{Zutic04} 
I. Zutic, 
J. Fabian, 
S. Das Sarma, 
{\em Spintronics: Fundamentals and applications}, 
Rev. Mod. Phys. {\bf 76}, 323-410 (2004).
%==============================================================================
% [10] MUST - citato all'inizio, Coulomb blockade
% ``Coulomb blockade of single-electron tunneling, and coherent oscillations in small tunnel juncitons''
\bibitem{Averin86} 
D.~V. Averin, 
K.~K. Likharev,
{\em Coulomb blockade of single-electron tunneling, and coherent oscillations in small tunnel junctions}, 
J. Low. Temp. Phys. {\bf 62}, 345-373 (1986).
%==============================================================================
% [11] MUST - citato all'inizio, spin blockade
% ``Current Rectification by Pauli Exclusion in a Weakly Coupled Double Quantum Dot System''
\bibitem{Ono02} 
K. Ono, 
D.~G. Austing, 
Y. Tokura, 
S. Tarucha, 
{\em Current Rectification by Pauli Exclusion in a Weakly Coupled Double Quantum Dot System}, 
Science {\bf 297}, 1313-1317 (2002).
%==============================================================================
% [12] MUST - citato all'inizio, double quantum dots
% ``Electron transport through double quantum dots''
\bibitem{vdWiel03} 
W.~G. van der Wiel, 
S. De Franceschi, 
J.~M. Elzerman, 
T. Fujisawa, 
S. Tarucha, 
L.~ P. Kouwenhoven, 
{\em Electron transport through double quantum dots}, 
Rev. Mod. Phys. {\bf 75}, 1-22 (2003).
%==============================================================================
% [13] citato all'inizio, room temperature
% ``Carbon nanotube single-electron transistors at room temperature''
\bibitem{Postma01} 
H.~W.~C. Postma, 
T. Teepen, 
Z. Yao, 
M. Grifoni, 
C. Dekker, 
{\em Carbon nanotube single-electron transistors at room temperature}, 
Science {\bf 293}, 76-79 (2001).
%==============================================================================
% [14] citato come all-electric QC - theo
% ``Pauli spin blockade in the presence of strong spin-orbit coupling''
\bibitem{Danon09} 
J. Danon, 
Yu.~V. Nazarov, 
{\em Pauli spin blockade in the presence of strong spin-orbit coupling}, 
Phys. Rev. B {\bf 80}, 041301R (2009).
%==============================================================================
% [15] citato come all-electric QC - Kouwen
% ``Spin-orbit qubit in a semiconductor nanowire''
\bibitem{Nadj10} 
S. Nadj-Perge, 
S.~M. Frolov, 
E.~P.~A.~M. Bakkers, 
L.~P. Kouwenhoven, 
{\em Spin-orbit qubit in a semiconductor nanowire}, 
Nature {\bf 468}, 1084-1087 (2010).
%==============================================================================
% [16] citato come all-electric QC - Kouwen
% ``Electrical control of single hole spins in nanowire quantum dots''
\bibitem{Pribiag13}
V.~S.Pribiag
S. Nadj-Perge,
S.~M.Frolov
J.~W.~G. van den Berg,
I. van Weperen
S.~R. Plissard,
E.~P.~A.~M. Bakkers
L.~P. Kouwenhoven
{\em Electrical control of single hole spins in nanowire quantum dots}
Nature Nanotech. {\bf 8}, 170-174 (2013).
%==============================================================================
% [17] citato come all-electric QC - Koppens
% ``Coherent Control of a Single Electron Spin with Electric Fields''
\bibitem{Nowack07} 
K.~C. Nowack, 
F.~H.~L. Koppens, 
Yu.~V. Nazarov, 
L.~M.~K. Vandresypen, 
{\em Coherent Control of a Single Electron Spin with Electric Fields}, 
Science {\bf 318}, 1430-1433 (2007).
%==============================================================================
% [18]  citato come all-electric QC
% ``Large spin-orbit coupling in carbon nanotubes''
\bibitem{SteeleNC13} 
G.~A. Steele
F. Pei,
E.~A. Laird, 
J.~M. Jol,
H.~B. Meerwaldt
L.~P. Kouwenhoven
{\em Large spin-orbit coupling in carbon nanotubes}
Nature Comm. {\bf 4}, 1573 (2013).
%==============================================================================
% [19]  citato come all-electric QC
% ``Tunable few-electron double quantum dots and Klein tunneling in ultraclean carbon nanotubes''
\bibitem{SteeleNN09} 
G.~A. Steele
G. Gotz
L.~P. Kouwenhoven
{\em Tunable few-electron double quantum dots and Klein tunneling in ultraclean carbon nanotubes}
Nature Nanotech. {\bf 4}, 363-367 (2009).
%==============================================================================
% [20] 
% ``Spin blockade and exchange in Coulomb-confined double quantum dots''
\bibitem{Weber14}
B. Weber
Y.~H. Matthias Tan
S. Mahapatra
T.~F. Watson,
H. Ryu,
R. Rahaman,
L.~C.~L. Hollenberg
G. Klimeck
M.~Y. Simmons,
{\em Spin blockade and exchange in Coulomb-confined double quantum dots},
Nature Nanotech. 91, 430-435 (2013).
%==============================================================================
% [21] InAs/InP DQD, non controllato davvero, no spin blockade
% ``Few Electron Double Quantum Dots in InAs/InP Nanowire Heterostructures''
\bibitem{Fuhrer07} 
A. Fuhrer, 
L.~E. Fr\"oberg, 
J.~N. Pedersen, 
M.~W. Larsson, 
A. Wacker, 
M.-E. Pistol, 
L. Samuelson, 
{\em Few Electron Double Quantum Dots in InAs/InP Nanowire Heterostructures}, 
Nano Lett. {\bf 7}, 243-246 (2007).
%==============================================================================
% [22] InAs/InP
% ``Few-Electron Quantum Dots in Nanowires''
\bibitem{Bjork04} 
M.~T. Bjork, 
C. Thelander, 
A.~E. Hansen, 
L.~E. Jensen, 
M.~W. Larsson, 
L.~R. Wallenberg, 
L. Samuelson, 
{\em Few-Electron Quantum Dots in Nanowires}, 
Nano Lett. {\bf 4}, 1621-1625 (2004).
%==============================================================================
% [23] InAs/InP
% ``Manipulation of Electron Orbitals in Hard-Wall InAs/InP Nanowire Quantum Dots''
\bibitem{Rod11}  
S. Roddaro, 
A. Pescaglini, 
D. Ercolani, 
L. Sorba, 
F. Beltram, 
{\em Manipulation of Electron Orbitals in Hard-Wall InAs/InP Nanowire Quantum Dots}, 
Nano Lett. {\bf 11}, 1695-1699 (2011).
%==============================================================================
% [24] InAs/InP
% ``Electrostatic Spin Control in InAs/InP Nanowire Quantum Dots''
\bibitem{Romeo12}  
L. Romeo, 
S. Roddaro, 
A. Pitanti, 
D. Ercolani, 
L. Sorba, 
F. Beltram, 
{\em Electrostatic Spin Control in InAs/InP Nanowire Quantum Dots}, 
Nano Lett. {\bf 12}, 4490-4494 (2012).
%==============================================================================
% [25] cotunneling di supporto per Fig.2
% ``Pauli spin blockade in cotunneling transport through a double quantum dot''
\bibitem{Liu2005} 
H.~W. Liu,
T. Fujisawa, 
. Hayashi, 
Y. Hirayama, 
{\em Pauli spin blockade in cotunneling transport through a double quantum dot}, 
Phys. Rev. B {\bf 72}, 161305(R) (2005).
%==============================================================================
% [26]
% ``Suppression of Spin Relaxation in an InAs Nanowire Double Quantum Dot''
\bibitem{Pfund07} 
A. Pfund, 
I. Shorubalko, 
K. Ensslin, 
R. Leturcq, 
{\em Suppression of Spin Relaxation in an InAs Nanowire Double Quantum Dot}, 
Phys. Rev. Lett. {\bf{99}}, 036801 (2007).
%==============================================================================
% [27]
% ``Phonon induced decay of the electorn spin in quantum dots''
\bibitem{Golovach04} 
V.~N. Golovach, 
A. Khaetskii, 
D. Loss, 
{\em Phonon induced decay of the electron spin in quantum dots}, 
Phys. Rev. Lett. {\bf 93}, 016601 (2004).
%==============================================================================
% [28]
\bibitem{Bellucci2004}
D. Bellucci,
M. Rontani,
F. Troiani,
G. Goldoni,
E. Molinari,
{\em Competing mechanisms for singlet-triplet transition in 
artificial molecules},
Phys. Rev. B {\bf 69,} 201308(R) (2004).
%==============================================================================
% [29]
{\modcol
\bibitem{Vdovin2000} 
E.~E. Vdovin,
A. Levin, 
A. Patan\`e, 
L. Eaves, 
P.~C. Main, 
Yu.~N. Khanin, 
Yu.~V. Dubrovskii, 
M. Henini, 
G. Hill 
{\em Imaging the electron wave function in self-assembled quantum dots }
Science {\bf 290}, 122-124 (2000).
}
%==============================================================================
% [30]
\bibitem{Panda2012} 
J.~K. Panda, 
A. Roy,
A. Singha, 
M. Gemmi, 
D. Ercolani, 
V. Pellegrini, 
L. Sorba
{\em Raman sensitivity to crystal structure in InAs nanowires}
Appl. Phys. Lett. {\bf 100}, 143101 (2012).

\end{thebibliography}
\end{document}

\author{Francesco Rossella}
\author{Andrea Bertoni}
\author{Daniele Ercolani}
\author{Massimo Rontani}
\author{Lucia Sorba}
\author{Fabio Beltram}
\author{Stefano Roddaro}

\title{SUPPLEMENTARY INFORMATION 

Nanoscale spin rectifiers controlled by the Stark effect}

\maketitle
\tableofcontents

\section{Device architecture}

\subsection{Nanowire properties and device structure}

\begin{figure}[h!]
\begin{center}
\ifshort
\includegraphics[width=0.8\textwidth]{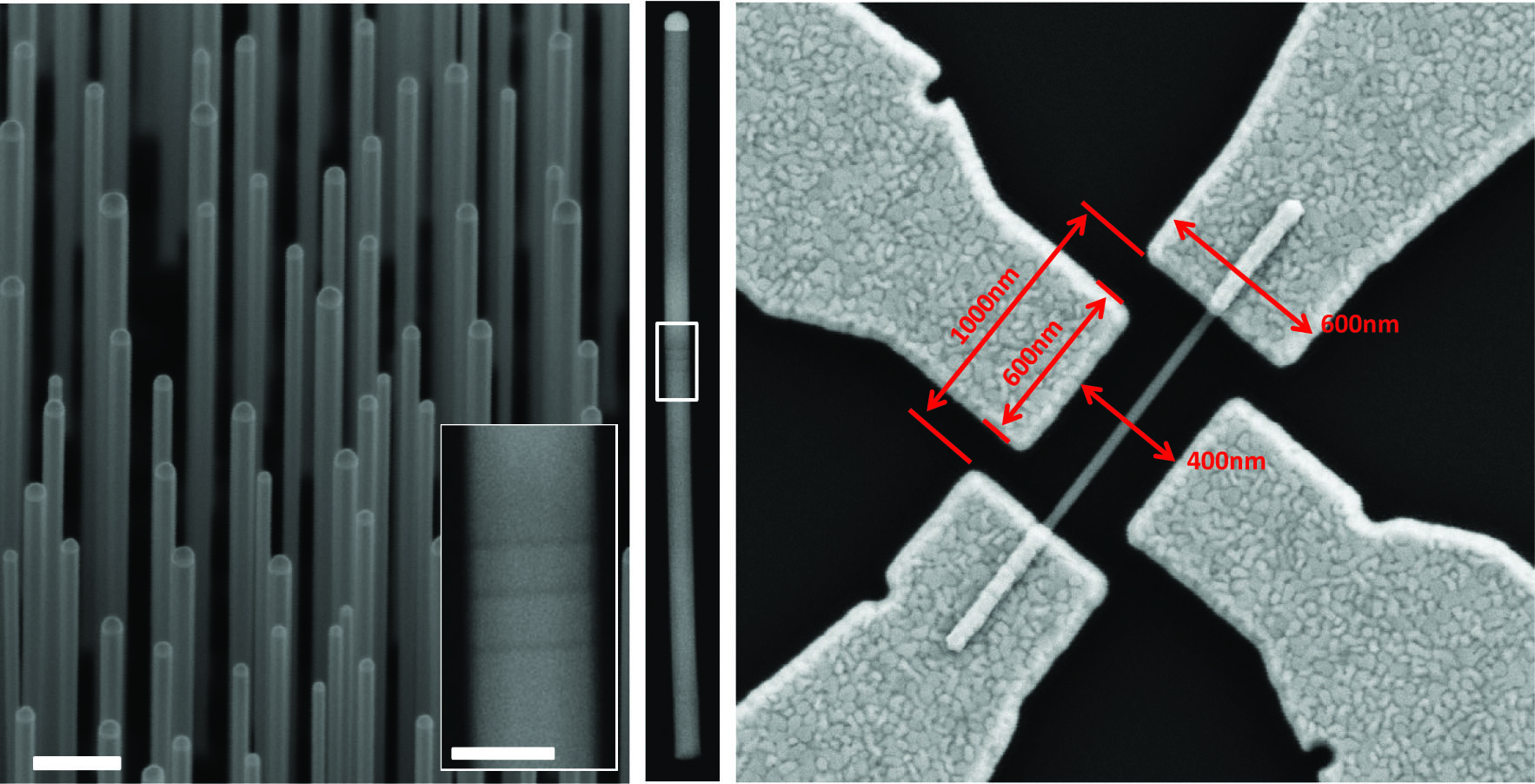}
\else
\includegraphics[width=0.8\textwidth]{Final_SF1.pdf}
\fi
\caption{From left to right: SEM picture of standing nanowires (scale bar $200\,{\rm nm}$); STEM image of a NW nominally identical to the one used in the experiment (inset scale bar $50\,{\rm nm}$); top view of one of the fabricated devices with quoted scales.}
\label{fig:Design}
\end{center}
\end{figure}

Details about the device structure are shown in Fig.~\ref{fig:Design}. On the left hand side we report a scanning electron microscope (SEM) image of as-grown standing nanowires. Nanowires were obtained by metal-seeded growth starting from Au dewetting on top of an InAs (111) substrate. Diameters were found to have a non-negligible statistical dispersion, with an average size of $55-60\,{\rm nm}$. Results reported in the main text were obtained on a device whose NW had a relatively large diameter of $70\,{\rm nm}$, based on SEM imaging performed at the end of the measurement session. NWs contained three $5\,{\rm nm}$ InP barriers defining two QDs with a nominal size of $20$ and $22.5\,{\rm nm}$. A scanning transmission electron microscope (STEM) picture is visible at the center of Fig.1, for a sample which is nominally identical to the one used in this study. STEM images were also used to determine the average distance between the heterostructure and the Au catalyst on the top of the NW with a $\approx100\,{\rm nm}$ precision: this information was crucial to align the gate layout to the heterostructures and to correctly place the DQD in between the lateral electrodes. 

Wires were transferred by drop-casting on a SiO$_2$/Si substrate (degenerately doped with a nominal resistivity of $1-5\,{\rm m\Omega\cdot cm}$). On the right hand side, we report an high-resolution picture of one of the fabricated devices with quoted dimensions.

\subsection{Multi-pole gating}
\label{SI_multipolegating}

Three independent gates are available in this architecture: two side ones ($SG1$ and $SG2$) visible in Fig.~\ref{fig:Design} and the degenerately-doped Si substrate ($bg$). In Fig.~\ref{fig:Stark} we report finite-element calculations showing the effect of these three gates on a dielectric body with hexagonal size and $\epsilon_r=15.15$. A $300\,{\rm nm}$ oxide with $\epsilon_r=3.9$ is also included while self-consistent carrier screening is here neglected. Focusing first on the simplified 2D version, a common biasing configuration ``$C$''~\cite{Nota1}, where all the gates are held to the same potential $V_{SG1}=V_{SG1}=V_{bg}=\nu$ will lead to a trivial shift of the potential in the hexagon with no significant electric field in any direction. In describing a general biasing configuration, two relevant further possibilities can be identified:

\begin{figure}[h!]
\begin{center}
\ifshort
\includegraphics[width=0.65\textwidth]{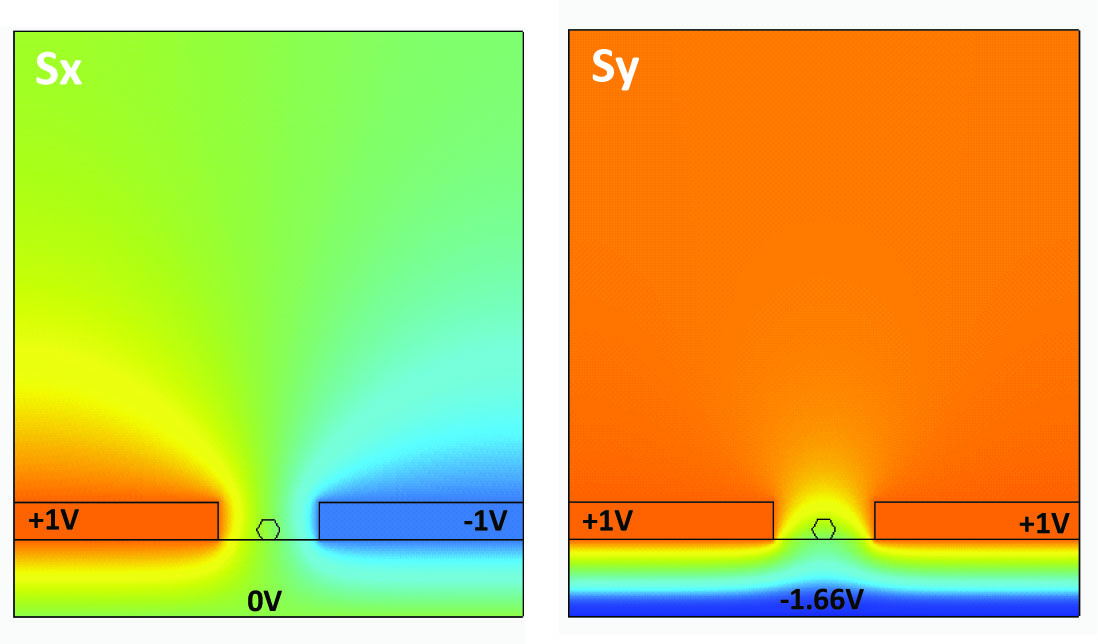}
\includegraphics[width=0.3\textwidth] {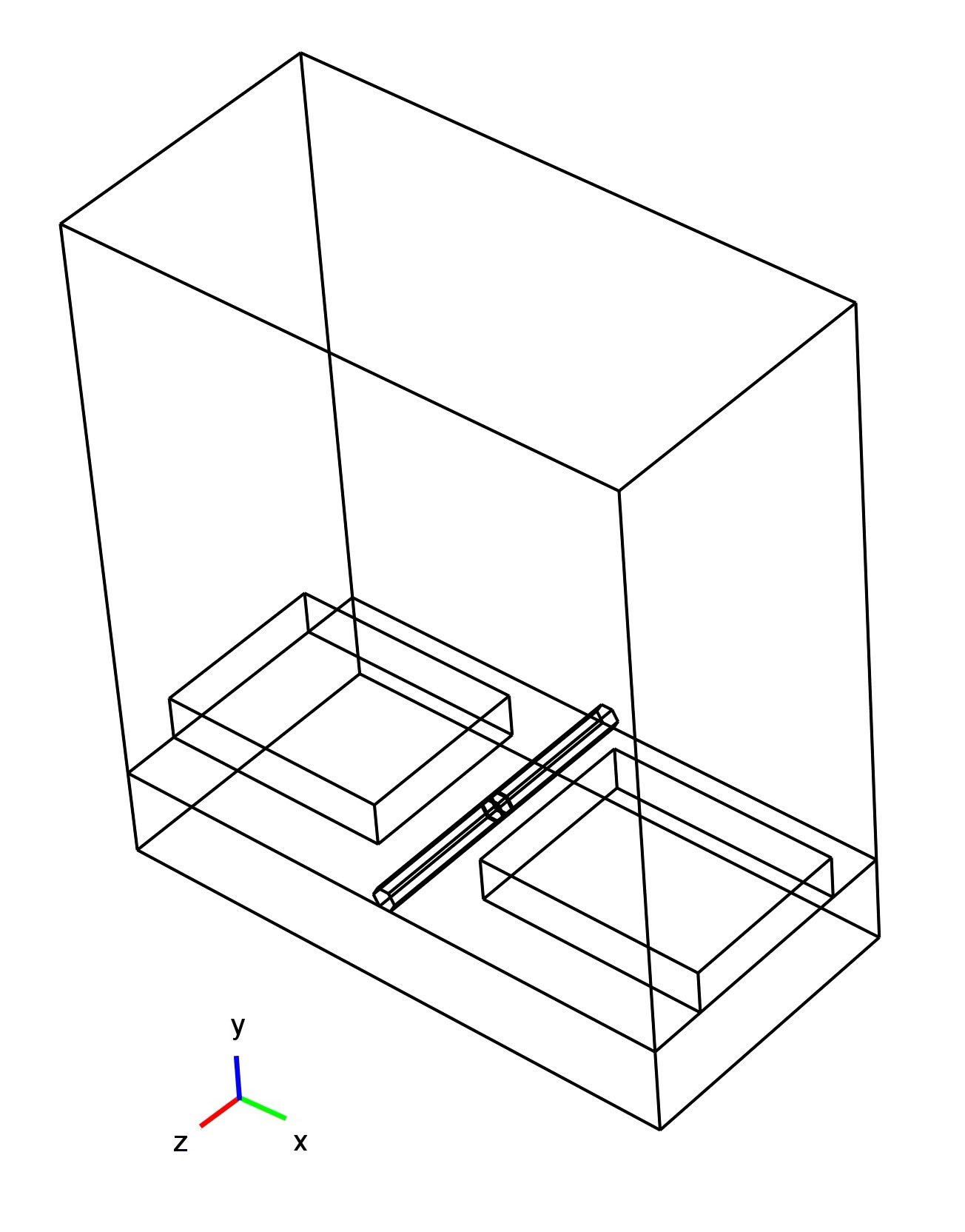}
\else
\includegraphics[width=0.65\textwidth]{Final_SF2ab.pdf}
\includegraphics[width=0.3\textwidth] {Final_SF2c.pdf}
\fi
\caption{Left: Gating configurations $S_x$ and $S_y$ inducing no potential shift and a finite transverse field in the $\hat{x}$ or $\hat{y}$ direction. Right: a 3D electrostatic simulation allow to estimate the screening effect due to the source and drain sections of the nanowire.}
\label{fig:Stark}
\end{center}
\end{figure}

\begin{itemize}
\item Configuration ``$S_x$'': $V_{SG1}=-V_{SG2}=\nu$ and $V_{bg}=0$, leading to a vanishing voltage shift in the hexagon and to an electric field oriented in the $\hat{x}$ direction;
\item Configuration ``$S_y$'': $V_{SG1}=V_{SG2}=\nu$ and $V_{bg}=-1.66\nu$, leading to a vanishing voltage shift in the hexagon and to an electric field oriented in the $\hat{y}$ direction.
\end{itemize}

Numerical calculation indicate that for ``$S_x$'' the expected field in the 2D case is $\langle E_x\rangle=\nu\times5.7\,{\rm kV/cm}$ with an average lateral spread of $\sqrt{\langle E_y\cdot E_y\rangle}=\nu\times1.2\,{\rm kV/cm}$. In the ``$S_y$'' the expected field in the 2D case is $\langle E_y\rangle=\nu\times9.3\,{\rm kV/cm}$ with an average lateral spread of $\sqrt{\langle E_x\cdot E_x\rangle}=\nu\times2\,{\rm kV/cm}$. Any general biasing configuration can be expressed as a linear superposition of the fundamental ones indicated here. {\modcol For instance, the experimental configuration studied in the main text corresponds to a lateral gate imbalance $\Delta V_{SG}=3-5\,{\rm V}$ which translates - in a purely $S_x$ configuration - into a lateral field of approximately $8.5-14.3\,{\rm kV/cm}$. A more refined estimate should also take into account the gate imbalance with respect to the backgate electrode. For instance, the configuration located at the center of Fig.2c of the main text corresponds to $V_{F}\approx-2.91\,{\rm V}$ and $V_{bg}\approx0.93\,{\rm V}$. For these values we obtain lateral gate voltages $V_{SG1}\approx-4.12\,{\rm V}$ and $V_{SG2}=0\,{\rm V}$. In turn this configuration can be written as a superposition of the $C$, $S_x$ and $S_y$ configurations and a final field $E_x\approx11.74\,{\rm kV/cm}$ and $E_y\approx10.45\,{\rm kV/cm}$ can be extracted.}

While this 2D simulation is quite useful to achieve a first-order estimate of the effect of a generic gating configuration, it should be stressed that these estimates can only be used to extract a rough order of magnitude of the actual impact on the dot electrostatic. The actual behavior of a real device will be impacted by gating asymmetries, screening by free charge in the QD and in the $S$ and $L$ sections of the NW. A 3D simulation was performed to estimate effect of the source and drain capacitive couplings, which are expected to play an important role even in an ideal InAs/InP device. These are in practice expected to screen the field induced on the QD region and tie its potential to the ground. The geometry used in the simulation is visible in the right panel of Fig.~\ref{fig:Stark} and included a perfect metallic screening by the $S$ and $D$ portions of the NW. In this worst case scenario we obtain a $\approx 50\%$ reduction in the average induced electric field in the QD region.

\section{Temperature dependence of the spin blockade}

\begin{figure}[h!]
\begin{center}
\ifshort
\includegraphics[width=0.6\textwidth]{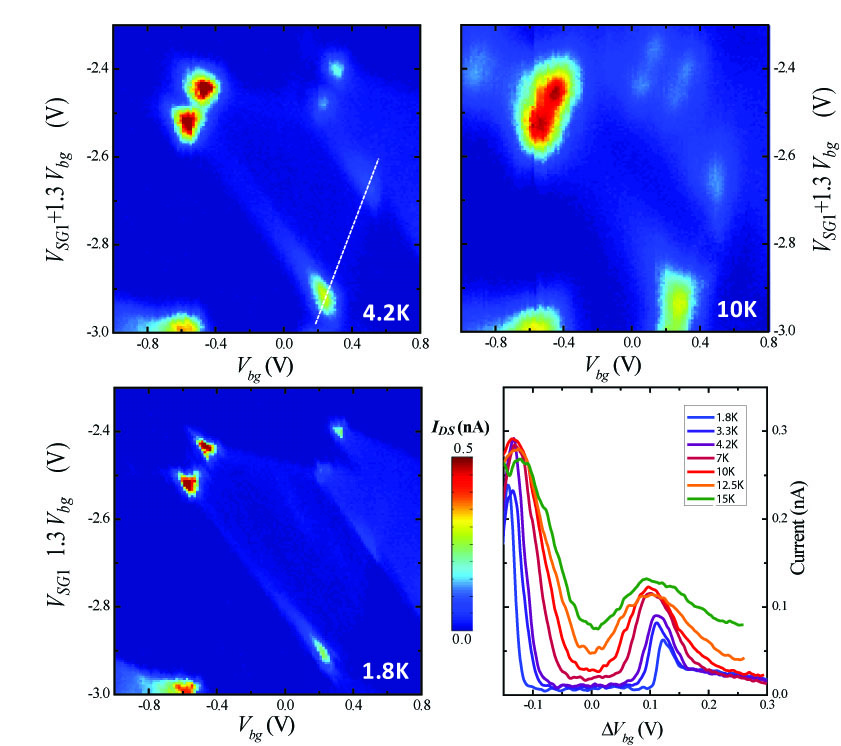}
\else
\includegraphics[width=0.6\textwidth]{Final_SF3.pdf}
\fi
\caption{A strong spin-blockade can be observed up to a temperature of the order of $10\,{\rm K}$. In the bottom right panel we report a cross-cut along the base of the various bias triangles as a function of $V_{bg}$. Gate voltages were shifted in order to position the zero at the middle between the two leakage peaks involving carrier exchange between the leads and the QD. {\modcol Measurements were performed at a bias $V_{DS}=-10\,{\rm mV}$.}}
\label{fig:SB}
\end{center}
\end{figure}

Provided the large singlet-triplet energy gap in our system ($\Delta_{ST}=9.1\pm0.3\,{\rm meV}$, see next sections) and charging energies, spin-blockade could be observed up to quite large temperatures and its evolution is visible in Fig.~\ref{fig:SB}. 

We report the full gate scans over the bias triangles for a temperature $T=1.8\,{\rm K}$, $4,2\,{\rm K}$ and $10\,{\rm K}$. The thermally activated spin blockade lifting is extracted by looking at a cross-section at the base of the bias triangles and plotted in the fourth panel of the figure: the two peaks on the left and right end of the spin-blocked base correspond to configurations where spin can be randomized by the interaction with the leads through the $(1,1)\leftrightarrows{\modcol(1,0)}$ and $(1,1)\leftrightarrows{\modcol(1,2)}$ substitutions, respectively. The backgate shift $\Delta V_{bg}$ was defined in such a way that the best blockade configuration is at zero shift. Lateral peaks can be taken as a reference for a non-blocking QD-lead alignment. Lifting of spin-blockade starts to be significant above $7\,{\rm K}$ but a evident modulation is still visible at the top explored temperature of $15\,{\rm K}$. It is useful to note that the major limiting factor in the current devices is not $\Delta_{ST}$ but rather the charging energy in our current implementation, which indicates the energy distance between the dot-lead resonances giving rise to the $(1,1)\leftrightarrows{\modcol(1,0)}$ and $(1,1)\leftrightarrows{\modcol(1,2)}$ spin substitution mechanisms. In fact, the evolution for $T>7\,{\rm K}$ is consistent with an activation energy equal to half the charging energy of the $D$ dot. For $T<7\,{\rm K}$ the leakage current decreases more slowly with temperature and is probably dominated by more fundamental lifting mechanisms such as spin-orbit coupling and interaction with nuclear spins. These results indicate in order to achieve higher operation temperatures it is crucial to adopt devices with a larger charging energy. In our device architecture, this implies that first of all smaller $S$ and $D$ barrier capacitances are needed. This might be achieved using either NWs with a smaller diameters and/or thicker InP barriers. 

\section{Quantum dot parameters and excited state data}

\subsection{Main resonances in finite bias transport up to $15\,{\rm meV}$}

\begin{figure}[h!]
\begin{center}
\ifshort
\includegraphics[width=0.8\textwidth]{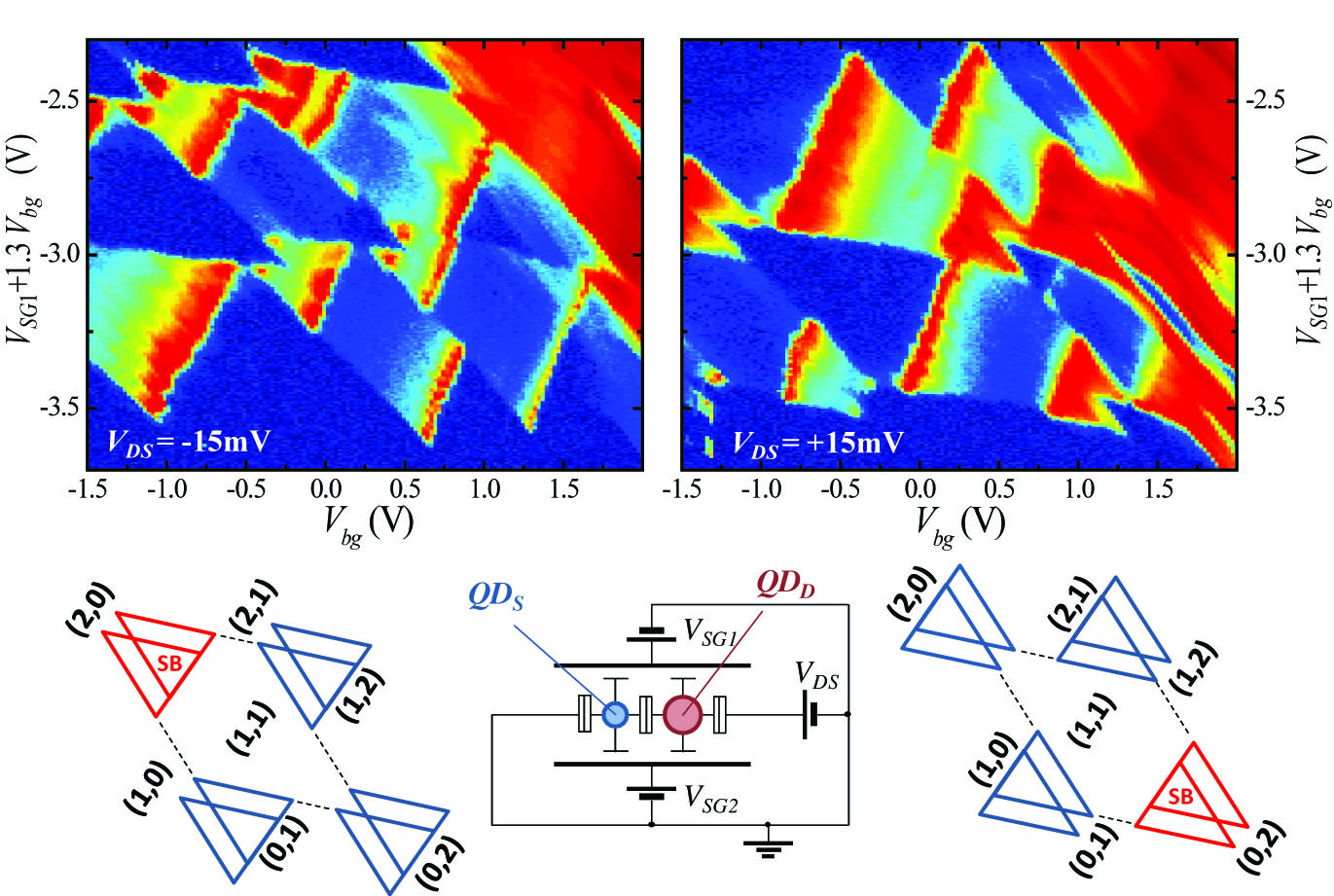}
\else
\includegraphics[width=0.8\textwidth]{Final_SF4.pdf}
\fi
\caption{Wide-range gate scans of the absolute current $|I_{DS}|$ for $V_{DS}=\pm 15\,{\rm mV}$ and $T=1.8\,{\rm K}$. In both bias directions the first available excited states are located at a large energy $\approx 10\,{\rm meV}$. }
\label{fig:Excited}
\end{center}
\end{figure}

\begin{table}[h!]
\begin{tabular}{c | c}
Bias & Tunnel configurations \& resonance energies\\ 
 & \\
\hline 
 & \\
$+15\,{\rm mV}\left\{\begin{array}{c c} & \\ & \\ \end{array}\right.$ &
$\begin{array}{c c c c} 
\hspace{0.2cm}(0,1) \to(1,0)\hspace{0.2cm} &
\hspace{0.2cm}(0,2) \to(1,1)\hspace{0.2cm} &
\hspace{0.2cm}(1,1) \to(2,0)\hspace{0.2cm} &
\hspace{0.2cm}(1,2) \to(2,1)\hspace{0.2cm} \\
10.1\,{\rm meV} & 10.7\,{\rm meV} & 9.1\,{\rm meV} & 9.9, 11.9\,{\rm meV} \\
\end{array}$ \\
 & \\
$-15\,{\rm mV}\left\{\begin{array}{c c} & \\ & \\ \end{array}\right.$ &
$\begin{array}{c c c c} 
\hspace{0.2cm}(0,1) \leftarrow (1,0)\hspace{0.2cm} &
\hspace{0.2cm}(0,2) \leftarrow (1,1)\hspace{0.2cm} &
\hspace{0.2cm}(1,1) \leftarrow (2,0)\hspace{0.2cm} &
\hspace{0.2cm}(1,2) \leftarrow (2,1)\hspace{0.2cm} \\
10.1\,{\rm meV} & \approx 10\,{\rm meV} & 10.3\,{\rm meV} & \approx 10\,{\rm meV}\\
\end{array}$ 
\end{tabular}
\caption{Observed excitation resonances below $15\,{\rm meV}$ for the various bias triangle pairs and for both the $V_{DS}$ signs. Data extracted from Fig.~\ref{fig:Excited}.}
\end{table}

Excitation spectrum of the studied device can be deduced from the large-scale/large-bias $|I_{DS}|$ scans we report in Fig.~\ref{fig:Excited}. The left and right panel refers to the $V_{DS}=\mp 15\,{\rm mV}$ case, respectively. In all the bias triangles, the first excited inter-dot resonance is located at $\approx 10\,{\rm meV}$ above the ground state, as a consequence of the strong confinement in our devices. Precise values for each couple of bias triangles and for both bias signs are indicated in Tab.~1. The errors were estimated as the peak half-width, which is consistent with thermal broadening of the conduction resonances (about $2k_BT\approx0.3\,{\rm meV}$ at $1.8\,{\rm K}$). The excitation gaps for the two triangle pairs on the right side of the colorplot at $V_{DS}=+15\,{\rm mV}$ can only be determined in a rough way due to a significant warping of the bias triangles. Nonetheless, they have obviously an amplitude which is comparable to the ones observed for the other filling configurations.

\subsection{Characteristic energies and gating parameters}

Other parameters can be extracted from the stability diagram, in particular the two QD have a charging energy of $E_{c,QD_D}=9\,{\rm meV}$ and $E_{c,QD_S}=15\,{\rm meV}$, respectively, with an estimated inter-dot coupling energy of $E_{c,m}=3\,{\rm meV}$. Following standard Coulomb blockade theory the corresponding capacitances~\cite{vdWiel03} are

\begin{eqnarray}
E_{c,QD_D} &=& e^2\frac{C_S}{C_DC_S-C_m^2} \\
E_{c,QD_S} &=& e^2\frac{C_D}{C_DC_S-C_m^2} \\
E_{c,m} &=& e^2\frac{C_m}{C_DC_S-C_m^2}.
\end{eqnarray}

This implies $C_S=11.4\,{\rm aF}$, $C_D=19.1\,{\rm aF}$ and $C_m=3.8\,{\rm aF}$. The values of $C_S$ and $C_D$ take into account various contributions to the dot charging energy and are expected to be dominated by the InP barrier capacitive coupling. {\modcol The obtained values are consistent with the actual explored geometry. It is also possible to define a set of {\em effective} lever arms for the gating parameters $V_{bg}$ and $V_F=V_{SG1}+1.3V_{bg}$ used in the main text colorplots:

\begin{eqnarray}
\alpha_{S,F}  &=& 29.3-33.0\,{\rm meV/V} \\
\alpha_{D,F}  &=& 21.1-24.8\,{\rm meV/V} \\
\alpha_{S,bg}   &=& 1.5-4.2\,{\rm meV/V} \\
\alpha_{D,bg}   &=& 10.1-15.5\,{\rm meV/V}
\end{eqnarray} 

\noindent where the relatively large range of values} takes into account the variation along the stability diagram and, in particular, the larger lever arms on the bottom right end of the scan. This is not very surprising given the relatively strong electric field effects expected for our device, which can easily give rise to non-linear effects in the evolution of the QD spectra as a function of the gate voltages.

\subsection{Excited states in the $(1,2)\to(2,1)$ bias triangle pair}

\begin{figure}[h!]
\begin{center}
\ifshort
\includegraphics[width=0.6\textwidth]{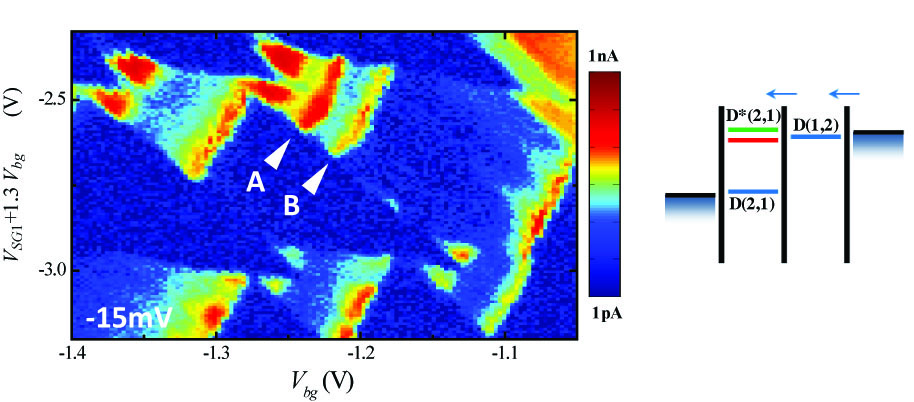}
\else
\includegraphics[width=0.6\textwidth]{Final_SF5.pdf}
\fi
\caption{Extract of the bias triangles at $V_{DS}=-15\,{\rm mV}$ and $T=1.8\,{\rm K}$, which presents two closely-spaced resonance lines at $\approx10\,{\rm meV}$. A possible interpretation involves two possible three-electron doublets as the final tunnel state with filling $(2,1)$.}
\label{fig:Excited2}
\end{center}
\end{figure}

It is worth to compare resonance lines in the $(1,2)\to(2,1)$ bias configuration at $V_{DS}=+15\,{\rm mV}$ with the other bias triangle couples for the same bias sign. Indeed, only in that case two closely-spaced resonances at $9.9\,{\rm meV}$ and $11.9\,{\rm meV}$ (see $A$ and $B$ marks in Fig.\ref{fig:Excited2}) are observed. If we make the reasonable working hypothesis that deeper filled levels do not play a significant role, the excitation spectrum in the $(0,1)\to(1,0)$ and $(0,2)\to(1,1)$ configurations is expected to be trivially related to the single-particle levels of the source-side QD ($QD_S$). The other two tunnel configurations are more subtle since they are both expected to involve final states with non-trivial spin configurations. This is obvious in the SB case of $(1,1)\to(2,0)$ where the electrons in the $QD_S$ can end up in an excited state which is compatible with a $T^*(2,0)$ spin triplet, which is responsible for the breakdown of SB. In addition, one can also expect a conduction resonance involving a $S^*(2,0)$ excited singlet configuration. It is unclear whether this conduction mode is not distinguishable from the triplet resonance or it is located at a significantly higher energy. Based on previous results~\cite{Romeo12}, the exchange gap between a singlet and triple state in our system could easily be of the order of various meV. The $(1,2)\to(2,1)$ case is apparently simpler, since no singlet-triplet gaps can be expected and both the initial and final states have to be a doublet due to spin conservation during tunneling. Nevertheless, two distinct resonances are clearly visible. Indeed, if the final excited $D^*(2,1)$ state involved two different orbitals in the left dot, one has three spins with general orientations. This means that in principle two doublet $D^*_1(2,1)$ and $D^*_2(2,1)$ and one quadruplet $Q^*(2,1)$ configuration are possible for the excited $(2,1)$ configuration. Quadruplet states can be excluded because of spin conservation, but two doublet configurations remain available and can give rise to two different excitation lines. These are likely to have distinct energies due to different exchange Coulomb interaction in the two $D^*_1$ and $D^*_2$ configurations. This phenomenology might thus indicate the realization of two distinct three-body doublets and implies non-trivial manybody correlations. While it is tempting to make such speculations, up to the present analysis this possibility could not be clearly verified and available scans in magnetic field are not conclusive yet.

\section{Numerical simulations of quantum states}

\subsection{Stark effect}

In the measurements described in the main text, we use the two side gates (see Fig.~1) to create an electrostatic potential gradient orthogonal to the NW axis.
This allows an external control of the energy levels of the two dots and eventually their alignment, within the transport window.
In order to confirm our hypothesis and assess the effect of a transverse electric field on the confined states of our DQD system we performed a set of numerical simulations where the effective-mass Schr\"odinger equation is solved in a 3D prismatic domain, representing the NW region within the source and drain leads.

\begin{figure}[h!]
\begin{center}
\ifshort
\includegraphics[width=0.4\textwidth]{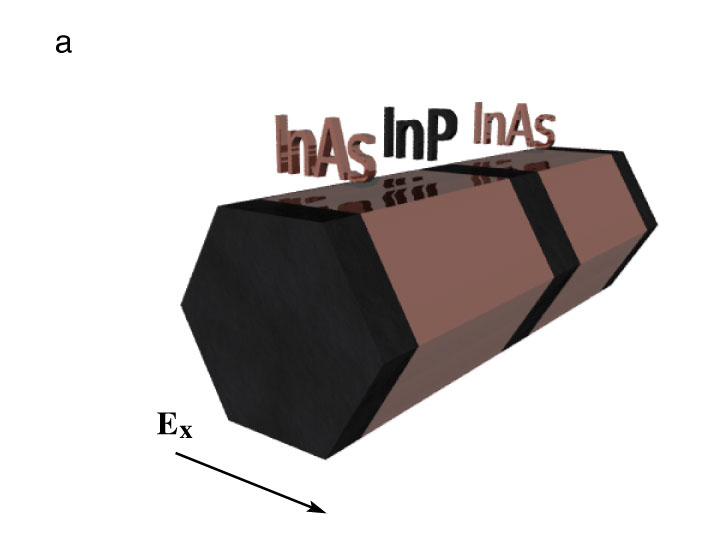}
\includegraphics[width=0.22\textwidth]{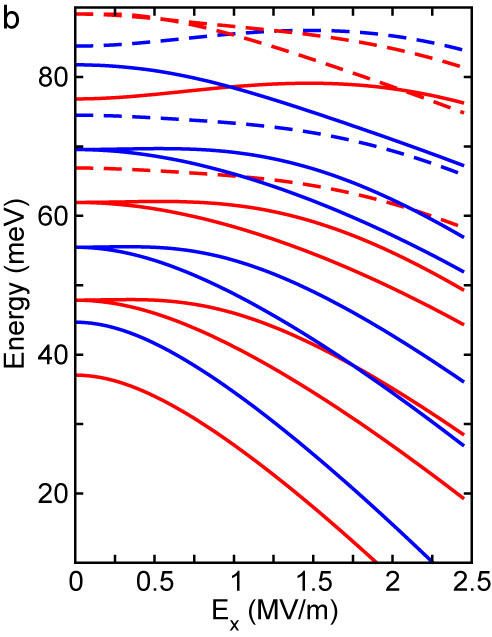}
\else
\includegraphics[width=0.4\textwidth]{Final_SF6a.pdf}
\includegraphics[width=0.22\textwidth]{Final_SF6b.pdf}
\fi
\caption{ (a) 3D domain of the numerical simulation assessing the effect of a uniform electric field in the $x$ direction, as indicated by the arrow. (b) Energy levels of the DQD mainly localized in the larger (red) or smaller (blue) InAs island vs the transverse electric field. Solid (dashed) curves represent states of the first (second) shell, with zero (one) closed-loop nodal line, as shown in Fig.~\ref{figSI_T2}. At $E_x=0$, the 1-2-2-1 degeneracy pattern typical of hexagonal systems is present.  At $E_x>0$ several level crossings occur since the Stark shift is different for orbitals with different symmetry.}
\label{figSI_T1}
\end{center}
\end{figure}

The simulation domain, shown in Fig.~\ref{figSI_T1}a, has an hexagonal section, with a maximal diameter of $80$~nm, while along the transport direction two InAs dots, $20$ and $22.5$~nm thick, are separated by a $5$~nm InP barrier. The effective mass of InAs and InP are taken $0.023$ and $0.080$, respectively, and a conduction-band offset of $600$~meV is used. An additional potential, increasing linearly along $x$ (Fig.~\ref{figSI_T1}a) is included, corresponding to an electric field from $E_x=0$ to $E_x=2.5\times 10^{6}$~V/m.

The simulation relies on a finite-volume method with a real-space regular hexagonal tessellation of the $x-y$ section and a linear discretization along $z$, leading to about $480\times10^3$ elementary volumes. Dirichlet conditions with $\psi=0$ are imposed at the boundaries for the wave function $\psi$. The confined eigenfunctions and energies are obtained by a Lanczos-type iterative approach\cite{arpackUG}. 

Results reported in Fig.~\ref{figSI_T1}b confirm that our conjecture on the Stark mechanism is valid. Before describing the details of the Stark spectrum, we want to stress that the present simulations should not be taken as a quantitative evaluation of the DQD levels. Rather, they give a rough estimate of the energies behavior and validate our model. Indeed, the fine details of the electric field induced by the three gates are not known inside the NW, as stated in Section~\ref{SI_multipolegating}, and the effect of surface states can have a remarkable effect on the DQD levels. Electrostatic simulations reported in Section~\ref{SI_multipolegating} indicate that the gating configuraton ``$S_x$'' gives an almost constant electric field in the NW area, thus we use a linear potential in our quantum simulations.

\begin{figure}[h!]
\begin{center}
\ifshort
\includegraphics[width=0.35\textwidth]{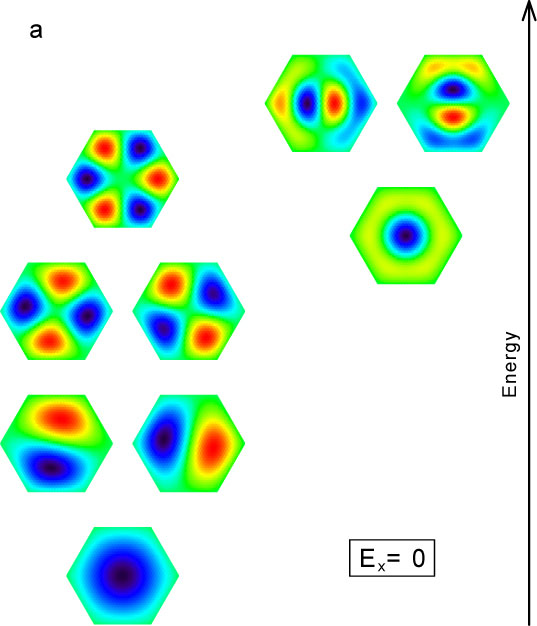}
\includegraphics[width=0.35\textwidth]{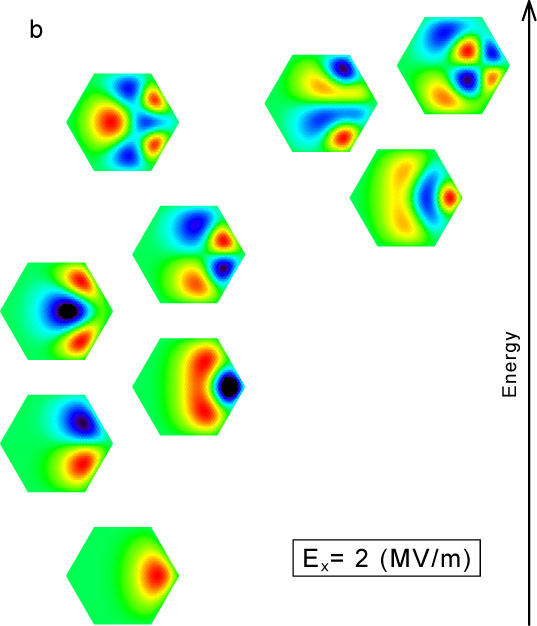}
\else
\includegraphics[width=0.35\textwidth]{Final_SF7a.pdf}
\includegraphics[width=0.35\textwidth]{Final_SF7b.pdf}
\fi
\caption{ {Section of the lowest nine wave functions taken in the middle (axis direction) of the large dot, without (a) and with (b) an electrostatic field applied in the horizontal direction. The states represented here correspond to the red lines of Fig.~\ref{figSI_T1}b, with the six states on the left side in each panel corresponding to solid lines, and the three on the right side corresponding to dashed lines and belonging to a different shell. Red, green and blue colors indicate positive, zero and negative values, respectively. The wave functions are arranged with increasing energy from the bottom to the top of each panel.
}}
\label{figSI_T2}
\end{center}
\end{figure}

\begin{figure}[h!]
\begin{center}
\ifshort
\includegraphics[width=0.6\textwidth]{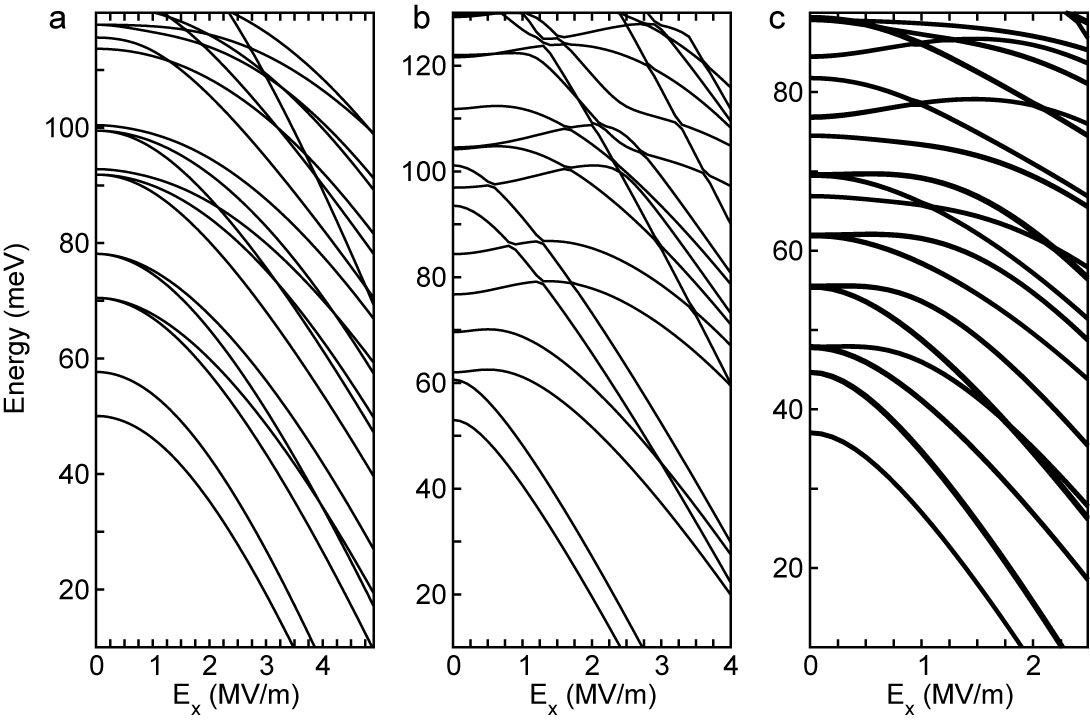}
\else
\includegraphics[width=0.6\textwidth]{Final_SF8.pdf}
\fi
\caption {\modcol Energy levels of the DQD vs the transverse electric field for the same system of Fig.~6b with (a) an additional 2D harmonic confinement in the plane of the hexagonal section with $\hbar\omega=20$~meV, (b) 1D harmonic confinement along $y$ direction with $\hbar\omega=40$~meV, (c) from four to eight repulsive scattering centers due to disorder with size of 2~nm and energy between 20 and 80~meV: here we superimpose eight simulations with a random position, energy and number of the impurities. The crossing of levels belonging to different dots is present in all the graphs and occurs at the same order of magnitude of the field.}
\label{figSI_TN1}
\end{center}
\end{figure}

In Fig.~\ref{figSI_T1}b we report the energy levels of the two dots against the transverse electric field $E_x$. Since the two dots are unequal, each state is completely localized in one of the two InAs regions and no delocalized level is found. In the ideal case where electrons in the two dots experience exactly the same radial confinement, this is true also for the field-induced degeneracy points: the levels that cross have a different symmetry in the $x-y$ hexagonal section and thus they do not mix.
We found the degeneracy pattern 1-2-2-1 typical of hexagonal structures\cite{EPL2013Ballester} for each of the dots, plus higher sets of orbitals with closed nodal lines around the center of the hexagon. For illustrative purposes we show in Fig.~\ref{figSI_T2} the $x-y$ wave function profiles of the first six states localized in the largest dot (red curves in Fig.~\ref{figSI_T1}b - those in the left dot are similar), at $E_x=0$ and $E_x=2\times 10^{6}$~V/m. The electrostatic field brakes the degeneracy and pushes the electron density towards a side. However, the character of the orbitals (as the number of lobes) is still clearly distinguishable.

The main features of the Stark spectrum of Fig.~\ref{figSI_T1}b are the energy level crossings, as the one around $E_x=1.7\times 10^{6}$~V/m between orbital $p_x$ of one dot and orbital $p_y$ of the other. As explained in the main text, this shows how the energy shift induced by a transverse field is different for different orbitals. {\modcol We also note that the investigated stability diagram of Fig.~2c, 3 and 4 in the main text occurs at a lateral gate imbalance of the order of $3-5\,{\rm V}$. Considering a spacing of $400\,{\rm nm}$ between the two side gates the field in the region of the NW can be approximately estimated to be $7.5-12.5\times 10^6 \,{\rm V/m}$. The actual field inside the NW is however expected to be smaller by a factor $\approx 10$ due to the actual 3D geometry of the device and to the large dielectric constant in InAs. A non-negligible vertical component should be expected too (see Section B in the Supplementary Information).  Overall estimated Stark field in our experimental configuration has thus a good match with the $1\times10^6\,{\rm V/m}$ scale obtained in our quantum mechanical simulations.

For the sake of completeness, we performed an extensive set of 3D simulations like the one presented in Fig.~\ref{figSI_T1}, including additional confining or disorder potentials, namely 

\begin{itemize}
  \item  2D harmonic confinement in the plane of the hexagonal section with $\hbar\omega$ from 5~meV to 80~meV, thus essentially changing the wave functions symmetry from prismatic to cylindrical and concentrating the electron probability in the NW core;
  \item  1D harmonic confinement along $y$ direction with $\hbar\omega$ from 5~meV to 80~meV, thus squeezing the wave functions about a plane parallel to the substrate;
  \item  fixed linear potential along $y$ direction, thus mimicking the effect of the back gate and pushing the wave functions towards it;
  \item  several configurations of localized impurities (from four to eight) with an extension of 2~nm and a positive or negative energy from 20 to 80~meV.
\end{itemize}

In all the above cases, the qualitative features of Fig.~\ref{figSI_T1}b are present and the Stark spectra are very similar, with the crossing of first and second dot levels induced by the field. This indicates that the control of level alignment with a transverse field is robust against disorder and symmetry breaking. As an example, we report in Fig.~\ref{figSI_TN1} the energy levels of the DQD vs the transverse electric field for 2D harmonic confinement, 1D harmonic confinement and disorder. Details are given in the caption.}

\begin{figure}[h!]
\begin{center}
\ifshort
\includegraphics[width=0.4\textwidth]{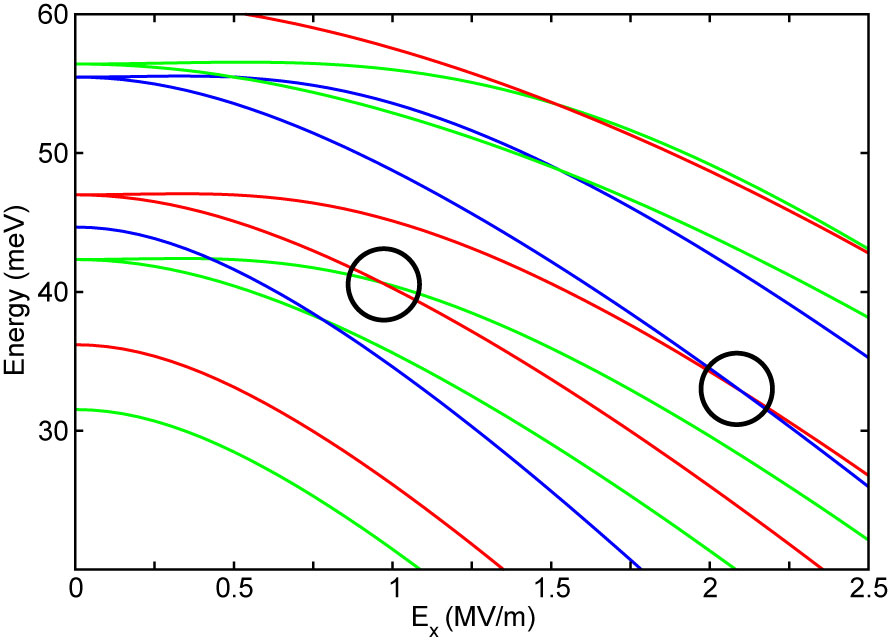}
\else
\includegraphics[width=0.4\textwidth]{Final_SF9.pdf}
\fi
\caption{ {\modcol Energy levels of a triple quantum dot system vs the transverse electric field. The first two dots have the same dimensions of the sample simulated in Fig.~\ref{figSI_T1}, namely 20 (blue curves) and 22.5 (red curves) nm, and the third dot (green curves) is 25 nm wide; the three dots are separated by 5 nm InP barriers, as in Fig. 6. The two black circles show the alignment of each couple of adjacent dots for a suitable value of the field.
}}
\label{grafRT.1}
\end{center}
\end{figure}

{\modcol

\subsection{Multi-dot systems}

A possible line of investigation opened by the present results consists in the study of the control of multi-dot systems without using local gates. In general, it is obviously {\em not} possible to achieve a full independent control of $n>2$ separate QDs since the Stark effect parameter $E_x$ does not provide sufficient degrees of freedom to individually address more than two dots at a time. On the other hand, the effect we describe could still allow a selective alignment of the levels located in each pair of adjacent dots, provided that Stark effect has a different impact on them. For instance, considering the specific NW implementation here discussed, dots should have different axial dimensions.

We numerically demonstrate this is indeed possible by simulating a device containing three dots with a thickness of 20, 22.5 and 25~nm (see Fig.~\ref{grafRT.1}). Since each pair of dots has a different axial thickness, the same concept demonstrated in our work makes it possible to align the levels of adjacent QDs (two of these conditions are marked by black circles in the figure) by tuning $E_x$.  In principle, this could allow inducing controlled single-hopping events along the QD chain. The method demonstrated in our work could for instance be possibly used to implement a single-electron turnstile that does not require a local gating architecture. Such a {\em partial} control could in fact be extended to an arbitrary number of dots, provided that the technological limitations imposed by the fine control of their dimensions during the growth process and the fine tuning of the gate potentials necessary to operate this kind of device architecture are overcome.}

\subsection{SB modulation with B }

Figure 5d in the main text shows bonding-antibonding energy splits of the first {\modcol five radial modes} of a DQD model system, as a function of the magnetic field. It has the purpose of exposing that a magnetic field indeed leads to an oscillatory pattern of the bonding-antibonding energy difference. In these simulations, the domain is two-dimensional (only the plane orthogonal to the magnetic field is considered), $140$~nm wide, with two $22.5$~nm thick InAs dots separated by a $5$~nm InP barrier. We use the same material parameters reported in the previous subsection.

The simulation is based on a 2D version of the finite-volume method, with the magnetic field included via Peierls substitutions in the Landau gauge, leading to factors $\exp{\left(i\frac{eB}{\hbar}y\right)}$ in the coupling term of two elements along $x$ direction~\cite{PR1969Langbein}.

\begin{figure}[h!]
\begin{center}
\ifshort
\includegraphics[width=0.5\textwidth]{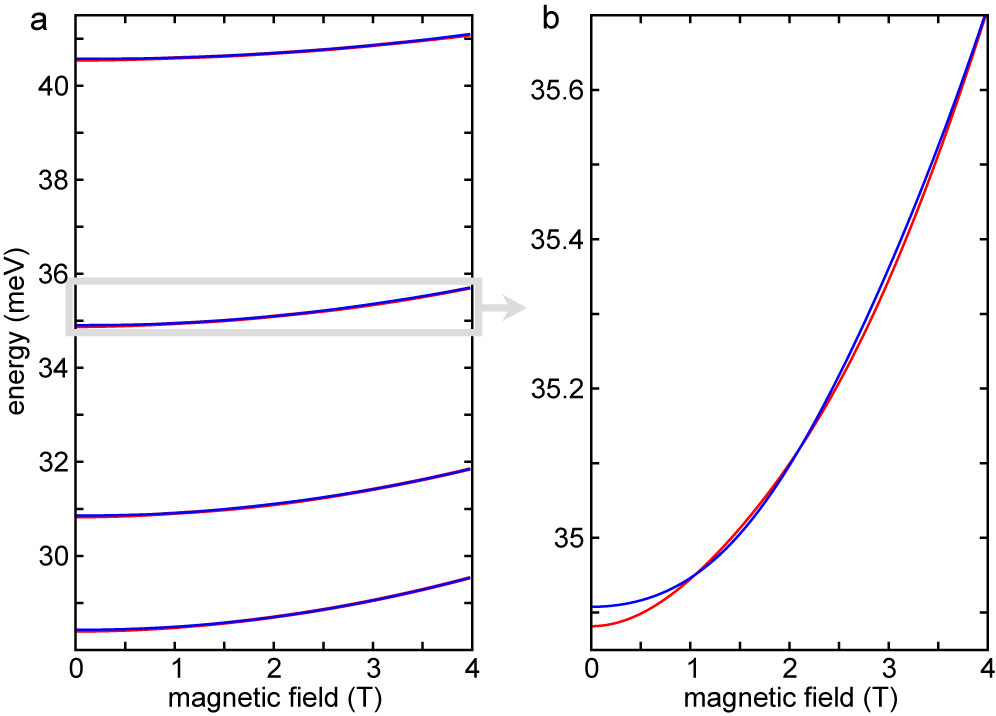}
\includegraphics[width=0.31\textwidth]{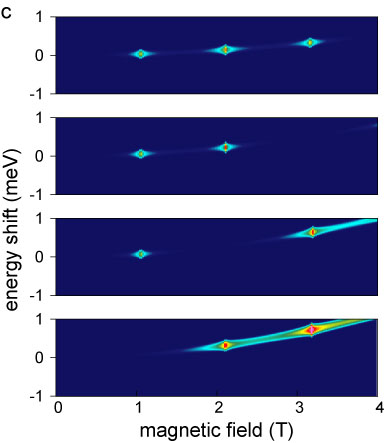}
\else
\includegraphics[width=0.5\textwidth]{Final_SF10ab.pdf}
\includegraphics[width=0.31\textwidth]{Final_SF10c.pdf}
\fi
\caption{ (a) Lowest eight energy levels of the 2D DQD model system described in the text, as a function of a transverse magnetic field. Due to the weak coupling, each bonding-antibonding couple is almost degenerate and appears as a single curve. (b) Detail of the spectrum showing the third couple. Here the multiple crossings of the two levels are distinguishable. The difference of the two energies vs $B$ is reported in Fig.~5d of the main text, for the first {\modcol five radial modes}.
(c) Theoretical intensity calculated with Eq.~(\ref{theointensity}) for the four couples of levels shown in (a), with increasing energy from the bottom to the top. Blue color indicates large bonding-antibonding split, thus an effective SB; red color indicates the neighborhood of a crossing, thus a SB lift. Colors are {\modcol in} logarithmic scale, following the palette reported in Fig.~\ref{figSI_T4}. The energy ``thickness'' of each stripe is induced by the temperature $T=1.7$~K corresponding to the experimental condition of Fig.~4b of the main text.}
\label{figSI_T3}
\end{center}
\end{figure}

A typical result of the above simulations is reported in Fig.~\ref{figSI_T3}a, where the first eight bonding (red) and antibonding (blue) levels are shown. The oscillations of the energy splits is very small and hardly distinguishable in the left panel (a) but it is clear in the detail of the right panel (b), where the third couple alone is shown.
The difference between the energies of each couple is reported in Fig.~5d of the main text.

In order to obtain a representation directly comparable with Fig.~4b of the main text and to include the effect of the finite temperature $T=1.7$~K on our simulation, we added a phenomenological thermal broadening using the typical form of the temperature-induced width of a Coulomb blockade peak\cite{beenakkerPRB44_1646}. Specifically, assuming that the effectiveness of the spin blockade is proportional to the bonding-antibonding split $\Delta$ (as explained in the main text), we computed, at every value of the magnetic field $B$ and for every bonding-antibonding couple, the corresponding theoretical intensity as
 
\begin{equation} \label{theointensity}
I(B,E)= I_0(B)\;\cosh^{-2}\left(\frac{\mathcal{E}(B)-E}{2K_B T}\right) ,
\end{equation}

where $K_B$ is the Boltzmann constant, $I_0(B)\propto(1-\Delta(B))$ is the maximal intensity for a given couple and is higher when the levels cross, $\mathcal{E}(B)$ is the mean energy of a bonding-antibonding couple. The results for the first four couples of states are reported in Fig.~\ref{figSI_T3}c, where the $y$ axis represents the magnetic energy shift, rather than the absolute energy. To better understand the above figures, one may think of the four color stripes as originating from the four double curves of Fig.~\ref{figSI_T3}a, broadened in energy (vertical axis) by the temperature and whose maximal intensity (color scale) is the value of the split.

\begin{figure}[h!]
\begin{center}
\ifshort
\includegraphics[width=0.6\textwidth]{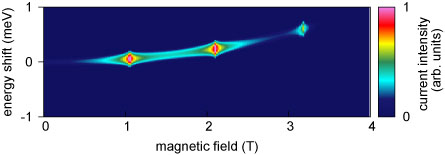}
\else
\includegraphics[width=0.6\textwidth]{Final_SF11.pdf}
\fi
\caption{ Average of the theoretical intensity (Eq.~\ref{theointensity}) for the first four couples (adding additional states does not change the result sigificatively), with $T=1.7$~K. Colors are in logarithmic scale. This plot should be compared with Fig.~4b of the main text. }
\label{figSI_T4}
\end{center}
\end{figure}

Following the same rationale explained in the main text for Fig.~5c, we compute an average of the above intensities and report it on Fig.~\ref{figSI_T4}. The intensity pattern obtained can be compared with the experimental data of Fig.4b, in the main text, reporting the current intensity as a function of the magnetic field applied and the side-gate voltage. In fact, the latter corresponds to a shift in the energy of the confined states, as in Fig.~\ref{figSI_T4}. Three common features of the two graphs can be stressed. First, a diamagnetic shift of the states. Second, a regular periodic SB lifting leading to three current peaks. Third, a strict SB condition for $B=0$ that is only loosely replicated at the following SB current minima. While the primary comparison with the experimental results should focus on the theoretical data of Fig.~5c of the main text (as the temperature broadening is included in Fig.~\ref{figSI_T4} at a phenomenological level), the similarities of the two plots strongly support our interpretation of the SB modulation with B. 
